\documentclass[a4paper,amsmath,amssymb,nofootinbib,showpacs,notitlepage,byrevtex,showkeys,prd%,preprint%,tightenlines%
]{revtex4-1}

\usepackage{graphicx}
\usepackage{subfigure}
\usepackage{bbm}

\newcommand{\Id}{ \mathbbm{1}}
\renewcommand{\vec}[1]{\boldsymbol{#1}}

\newcommand{\beq}{\begin{equation}}
\newcommand{\eeq}{\end{equation}}
\newcommand*{\deltabar}{\delta\mkern-8mu\mathchar'26}
\newcommand{\ii}{\mathrm{i}}
\newcommand{\dd}{\mathrm{d}}
\newcommand*{\da}[1][]{\mathop{\mathrm{d}\mkern-7mu\mathchar'26\mkern-1mu^{#1}}\mkern-4mu}
\newcommand{\Dc}{\mathcal{D}}
\newcommand{\Hc}{\mathcal{H}}
\newcommand{\Ic}{\mathcal{I}}
\newcommand{\Nc}{\mathcal{N}}

\newcommand{\tr}{\mathrm{tr}}
\newcommand{\Ncc}{N_{\mathrm{C}}}

\newcommand{\sinc}{\mathrm{sinc}}

\newcommand{\ve}{\vec{e}}
\newcommand{\vx}{\vec{x}}
\newcommand{\vy}{\vec{y}}
\newcommand{\vz}{\vec{z}}
\newcommand{\vp}{\vec{p}}
\newcommand{\vq}{\vec{q}}

\newcommand{\vA}{\vec{A}}
\newcommand{\vB}{\vec{B}}
\newcommand{\vD}{\vec{D}}
\newcommand{\vPi}{\mathbf{\Pi}}
\newcommand{\valpha}{\boldsymbol{\alpha}}

\newcommand{\hp}{\hat{\vp}}
\newcommand{\hq}{\hat{\vq}}

\newcommand{\hpq}{\widehat{(\vp + \vq)}}

\newcommand{\nf}{\mathfrak{n}}
\newcommand{\mf}{\mathfrak{m}}

\newcommand{\he}{\hat{\ve}}

\usepackage[font=small]{caption}
\begin{document}

\title{Chiral and deconfinement phase transition in the Hamiltonian approach to QCD in Coulomb gauge}

\author{H.~Reinhardt and P.~Vastag}
\affiliation{Institut f\"ur Theoretische Physik\\
Eberhard-Karls-Universit\"at T\"ubingen \\
Auf der Morgenstelle 14\\
D-72076 T\"ubingen\\
Germany}
\date{\today}

\begin{abstract}
The chiral and deconfinement phase transitions are investigated within the variational Hamiltonian approach to QCD in Coulomb gauge. The temperature $\beta^{- 1}$ is introduced by compactifying a spatial dimension. Thereby the whole temperature dependence is encoded in the vacuum state on the spatial manifold $\mathbb{R}^2 \times S^1 (\beta)$. The chiral quark condensate and the dual quark condensate (dressed Polyakov loop) are calculated as function of the temperature. From their inflection points the pseudo-critical temperatures for the chiral and deconfinement crossover transitions are determined. Using the zero-temperature quark and gluon propagators obtained within the variational approach as input, we find  $168 \, \mathrm{MeV}$ and $196 \, \mathrm{MeV}$, respectively, for the chiral and deconfinement transition.
\end{abstract}
\maketitle

\section{Introduction}

Understanding the QCD phase diagram is still one of the major challenges of particle physics. At low temperature and baryon density quarks and gluons are confined inside hadrons and chiral symmetry is spontaneously broken. When nuclear matter is heated up or strongly compressed it undergoes a phase transition to a plasma of correlated quarks and gluons. This deconfinement phase transition is accompanied by a restoration of chiral symmetry. 

The phase diagram of nuclear matter is intensively studied from both the experimental and theoretical side. Much insight into the temperature behavior of QCD has been obtained through lattice studies \cite{Karsch2002}. The lattice calculations fail, however, to treat QCD at a finite chemical potential due to the notorious sign problem \cite*{[{see e.g.~}][{and references therein}]Gattringer2016}: The quark determinant becomes complex for finite chemical potentials and gauge groups $SU(\Ncc > 2)$. Several methods have been invented to circumvent this problem. However, so far all of these methods are restricted to small chemical potentials. On the other hand, continuum approaches do not face this problem and can deal with large real chemical potentials. The description of the deconfinement phase transition as well as the description of the confinement phenomenon itself requires a non-perturbative treatment of QCD. During the last two decades non-perturbative continuum approaches have been developed and applied to the QCD vacuum and the deconfinement phase transition. These continuum approaches are based on either Dyson--Schwinger equations in Landau \cite{Fischer2006, *Alkofer2001, *Binosi2009} and Coulomb gauge \cite{Watson2006, *Watson2007, *Watson2008}, functional renormalization group flow equations in Landau gauge \cite{Pawlowski2007, *Gies2012} or variational calculations within the Hamiltonian formulation in Coulomb gauge \cite{Feuchter2004, *Feuchter2004a, *Feuchter2005, *ERS2007}. All three types of approaches are related to each other\footnote{For example, in a certain approximation the flow equations become Dyson--Schwinger equations. The latter are exploited in the variational Hamiltonian approach in order to deal with non-Gaussian wave functionals \cite{CR2010}.} and require at some stage certain approximations in order to be feasible. By comparing the results obtained in the different approaches their predictive power is increased.

The deconfinement phase transition is the transition from a center symmetric low temperature phase to a high temperature phase with center symmetry broken \cite{Svetitsky1986}. Therefore, each quantity which transforms non-trivially under center transformations can serve as an order parameter of confinement. The most prominent example is the expectation value of the Polyakov loop \cite{Svetitsky1986}. This quantity was studied in various continuum approaches \cite{BGP2010, MP2008, BH2012, Reinhardt2012, *RH2013, Fischer2009, FM2009, FMM2010, [][{and references therein}]{RSTW2016}, QR2016, Canfora2015} (see also ref.~\cite{Dumitru2012} for the study of the deconfinement phase transition in a random matrix model). In ref.~\cite{Gattringer2006} a remarkable relation between the Polyakov loop and the so-called dual condensate has been established, see also ref.~\cite{Synatschke2008}. In the usual functional integral approach, where the finite temperature $\beta^{-1}$ is introduced by compactifying the Euclidean time axis, the dual condensates $\Sigma_n$ (with $n$ integer) are defined by
\beq
\Sigma_n = \int\limits_0^{2 \pi} \frac{\dd \varphi}{2 \pi} \exp(-\ii n \varphi) \langle \bar{\psi} \psi \rangle_\varphi \, , \label{Gl: DualesKondensat}
\eeq
where the subscript $\varphi$ indicates that the quark fields $\psi$ satisfy the $U(1)$-valued boundary condition
\beq
\psi(x_4 + \beta/2, \vx) = \mathrm{e}^{\ii \varphi} \psi(x_4 - \beta/2, \vx) \label{Gl: RBphi}
\eeq
on the compactified Euclidean time axis. Within the lattice formulation of QCD it is not difficult to show that $\Sigma_n$ represents (the expectation value of) the sum of all closed Wilson loops winding $n$-times around the compactified Euclidean time axis. Thus, $\Sigma_1$ contains, in particular, the Polyakov loop together with all other Wilson loops that wind once around the compactified time axis. For this reason, $\Sigma_1$ was called ``dressed Polyakov loop''. It obeys the same behavior under center transformations as the Polyakov loop and can thus serve as an order parameter for the deconfinement phase transition.

The dual condensate was calculated on the lattice, see e.g.~ref.~\cite{Zhang2011}, and also in the Dyson--Schwinger approach in Landau gauge \cite{Pawlowski2007, *Gies2012} using quenched lattice data for the temperature dependent gluon propagator as input. In the present paper, we calculate the dual quark condensate (\ref{Gl: DualesKondensat}) as well as the ordinary chiral quark condensate $\langle \bar{\psi} \psi \rangle_{\varphi = \pi}$ as function of the temperature within the variational approach to QCD developed in ref.~\cite{Feuchter2004, *Feuchter2004a, *Feuchter2005, *ERS2007} for pure Yang--Mills theory and extended in refs.~\cite{Pak2013, *Pak2012a, QCDT0, QCDT0Rev} to full QCD. In the usual Hamiltonian formulation of a gauge theory, which is based on the canonical quantization in Weyl gauge $A_0 = 0$, neither the Polyakov loop nor the dual condensate can be evaluated. There the finite temperature is not introduced by the compactification of the time axis but the grand canonical partition function is obtained from the trace of the density operator $\exp [-\beta (H - \mu N)]$ in Fock space. Here $H$ and $N$ are the Hamiltonian and the particle number operator, respectively, and $\mu$ is the chemical potential. Such an approach to finite temperature Yang--Mills theory was pursued in ref.~\cite{Reinhardt2011, *Heffner2012} by making a quasi-particle ansatz for the density operator of the grand canonical ensemble and determining the quasi-gluon energies by minimizing the free energy. 

In ref.~\cite{Reinhardt2016}, an alternative Hamiltonian approach to finite temperature quantum field theory was proposed where the temperature is 
introduced by compactifying a spatial dimension. This approach is advantageous since it does not require the introduction of a statistical density operator. Instead, the whole temperature behavior is encoded in the vacuum state calculated on the spatial manifold $\mathbb{R}^2 \times S^1(\beta)$. Within this approach Yang--Mills theory was treated at finite temperature in ref.~\cite{Heffner2015} and results consistent with those of the grand canonical ensemble, ref.~\cite{Reinhardt2011, *Heffner2012}, were obtained.  Furthermore, in ref.~\cite{Reinhardt2012, *RH2013} the Polyakov loop was calculated within this approach for Yang--Mills theory. In the present paper, we apply this approach to QCD. We will calculate the 
temperature dependence of the order parameters of chiral symmetry breaking and confinement: the quark condensate and the dual quark condensate.

The organization of the rest of the paper is as follows: In section \ref{secII}, we briefly review the Hamiltonian approach to finite temperature quantum field theory introduced in ref.~\cite{Reinhardt2016}. In section \ref{sectIII}, we show how the dual quark condensate can be calculated within this approach. In section \ref{sectIV}, the Hamiltonian approach to QCD in Coulomb gauge is developed on the partially compactified spatial manifold $\mathbb{R}^2 \times S^1(\beta)$. The gauge fixed QCD Hamiltonian is given in subsection \ref{subsecA}, the variational ansatz for the QCD vacuum wave functional is presented in subsection \ref{subsecB} and the variational equations of the quark sector are given in subsection \ref{subsecC}. The expressions for the chiral and dual quark condensate within the Hamiltonian approach of section \ref{sectIV} are derived in section \ref{sectV}. Section \ref{Abschn: Numerik} contains our numerical results. Finally, our conclusions are given in section \ref{sectVII}. Some mathematical details are presented in appendices.

\section{Finite temperature from compactifying a spatial dimension}\label{secII}

Below, we briefly summarize the essential ingredients of the Hamiltonian approach to finite temperature quantum field theory by compactifying a spatial dimension, for which we choose the 3-axis \cite{Reinhardt2016}. For later use, we will present this approach immediately for the case of QCD. Let
\beq
\label{218-f1}
H(\beta) = \int_{\beta} \dd^3 x \, \Hc[\psi, A]
\eeq
denote the Hamiltonian of QCD on the partially compactified spatial manifold $\mathbb{R}^2 \times S^1(\beta)$ where the integration measure is given by
\beq
\int_{\beta} \dd^3 x := \int \dd x_1 \int \dd x_2 \int_{-\beta/2}^{\beta/2} \dd x_3 \, .
\eeq
Here, $\Hc[\psi, A]$ is the Hamiltonian density of QCD arising after canonical quantization in Weyl gauge $A_0 = 0$ and possibly further gauge fixing. 
On the spatial manifold $\mathbb{R}^2 \times S^1(\beta)$, the gauge field $\vA (\vx)$ and the quark field $\psi(\vx)$, respectively, satisfy periodic and anti-periodic boundary conditions in the compactified spatial dimension
\begin{subequations}
\label{Gl: RB}
\begin{align}
\vA(\vx_\perp, x_3 = \beta/2) &= \vA(\vx_\perp, x_3 = -\beta/2) \, , \\
\psi(\vx_\perp, x_3 = \beta/2) &= -\psi(\vx_\perp, x_3 = -\beta/2) \, .
\end{align}
\end{subequations}
As shown in ref.~\cite{Reinhardt2016}, the whole finite temperature properties can be extracted from the vacuum state on the partially compactified spatial manifold $\mathbb{R}^2 \times S^1 (\beta)$ with $\beta$ being the inverse temperature. To be more precise, the grand canonical partition function at temperature $\beta^{- 1}$ and chemical potential $\mu$ is given by
\beq
\label{237-f1-2}
Z(\beta, \mu) = \lim\limits_{l \to \infty} \exp\Bigl(-l \widetilde{E}_0(\beta, \mu)\Bigr) \, , 
\eeq
where $l \to \infty$ is the length of the uncompactified spatial dimensions and $\widetilde{E}_0(\beta, \mu)$ is the ground state energy of the modified Hamiltonian\footnote{Due to the interchange of the temporal axis with a spatial one the grand canonical partition function is given by
\[
Z(\beta, \mu) = \lim_{l \to \infty} \tr\exp\Bigl(-l \widetilde{H}(\beta, \mu)\Bigr) = \lim_{l \to \infty} \sum_n \exp\Bigl(-l \widetilde{E}_n(\beta, \mu)\Bigr)
\]
where the trace is over the whole Fock space so that the sum runs here over all eigenstates of $\widetilde{H}$ [Eq.~(7)]. From this sum in the limit $l \to \infty$ only the ground state survives.}
\beq
\label{242-*fa-3}
\widetilde{H}(\beta, \mu) = H(\beta) + \ii \mu \int_{\beta} \dd^3 x \, \psi^{\dagger}(\vx) \alpha_3 \psi(\vx) \, .
\eeq
Here $\alpha_3$ is the Dirac matrix corresponding to the compactified (spatial) dimension. Separating from the ground state energy $\widetilde{E}_0(\beta, \mu)$ the volume $l^2 \beta$ of the spatial manifold $\mathbb{R}^2 \times S^1(\beta)$,
\beq
\label{252-f1-4}
\widetilde{E}_0(\beta, \mu) = l^2 \beta e(\beta, \mu) \, ,
\eeq
one finds for the pressure
\beq
\label{257-f1-5}
P = -e(\beta, \mu)
\eeq
and for the thermal energy density
\beq
\label{262-f1-6}
\varepsilon = \frac{\partial}{\partial \beta} (\beta e) - \mu \frac{\partial e}{\partial \mu} \, .
\eeq
Let us stress that the above summarized approach is completely equivalent to the usual grand canonical ensemble as long as the relativistic invariance is preserved. It is, however, advantageous in non-perturbative investigations since it requires only the calculation of the ground state energy density on the spatial manifold $\mathbb{R}^2 \times S^1(\beta)$ but avoids the explicit treatment of the grand canonical density operator $\exp[-\beta (H - \mu N)]$. 

Due to the periodic and anti-periodic boundary conditions (\ref{Gl: RB}) of the fields, the third component of the momentum variable is discrete and given by the bosonic and fermionic, respectively, Matsubara frequencies
\beq
\label{287-f2}
\omega_{\nf} = \frac{2 \pi \nf}{\beta} \, , \quad \quad \Omega_{\nf} = \frac{2 \nf + 1}{\beta} = \omega_{\nf} + \frac{\pi}{\beta} \, .
\eeq
Furthermore, some mathematical manipulations are required to obtain meaningful results for thermodynamic quantities:
\begin{itemize}

\item[i)] The Matsubara sums have to be Poisson resummed using
\beq
\frac{1}{\beta} \sum_{\nf = -\infty}^{\infty} f(\omega_{\nf}) = \int\limits_{-\infty}^{\infty} \da p_3 \sum_{l = - \infty}^{\infty} \exp(\ii l \beta p_3) f(p_3) \, , \label{Gl: PoissonSummenformel}
\eeq
where $\da = \dd / (2 \pi)$. Here the integration variable on the r.h.s.~has been labeled as third component of the 3-momentum. This has the advantage that a Fourier integral for the partially compactified spatial manifold $\mathbb{R}^2 \times S^1(\beta)$ can be compactly written as
\beq
\int_{\beta} \da^3 p \, f(\vp_\perp, \omega_{\nf}) := \int \da^2 p_\perp \, \frac{1}{\beta} \sum_{\nf = - \infty}^{\infty} f(\vp_\perp, \omega_{\nf}) = \int \da^3 p \, f(\vp) \sum_{l = - \infty}^{\infty} \exp(\ii l \beta p_3) \, , \label{305-f2a-x1}
\eeq
where on the r.h.s.~we have the usual integration measure of the momentum space for a flat spatial manifold. From this representation it is clear that the $l = 0$ term in eq.~(\ref{305-f2a-x1}) represents the zero temperature (i.e.~$\beta \to \infty$) contribution. Thus, Poisson resummation allows an easy separation of the (usually divergent) $T = 0$ contributions to the various thermodynamic quantities.

\item[ii)] In the presence of a real chemical potential an analytic continuation of the chemical potential is required \cite{Reinhardt2016}. This is necessary in order to make the remaining Poisson sum convergent.

\end{itemize}

\section{The dual quark condensate} \label{sectIII}

In the Hamiltonian approach to finite temperature quantum field theory summarized in section \ref{secII}, where the original Euclidean time axis has become the compactified spatial $3$-axis, the quark fields in the dual condensate (\ref{Gl: DualesKondensat}) satisfy the $U(1)$-valued boundary condition
\beq
\label{680-3}
\psi(\vx_{\perp}, x_3 + \beta/2) = \mathrm{e}^{\ii \varphi} \psi(\vx_{\perp}, x_3 - \beta/2) \, .
\eeq
These fields $\psi$ can be related to fermion fields $\widetilde{\psi}$ satisfying the usual anti-periodic boundary conditions (\ref{Gl: RB}) by the $x_3$-dependent phase transformation
\beq
\label{686-4}
\psi(\vx) = \exp\left(\ii \frac{\varphi - \pi}{\beta} x_3\right) \widetilde{\psi}(\vx) \, .
\eeq
This phase can be absorbed into an imaginary chemical potential, c.f.~eq.~(\ref{242-*fa-3}). Indeed, for the Dirac Hamiltonian of the quarks with bare mass $m_{\mathrm{Q}}$ interacting with the gluon field $\vA = \vA^a t_a$ ($t_a$ denotes the generator of the color group in the fundamental representation),
\beq
H_{\mathrm{Q}}[\psi, A; \beta] = \int_{\beta} \dd^3 x \, \psi^\dagger(\vx) \Bigl[\vec{\alpha} \cdot \bigl(-\ii \nabla + g \vA(\vx)\bigr) + \gamma_0 m_{\mathrm{Q}} \Bigr] \psi(\vx) \equiv H_{\mathrm{Q}}^0 + H_{\mathrm{Q}}^A \label{Gl: DiracHamiltonian}
\eeq
where $\valpha$ and $\gamma_0$ are the usual Dirac matrices and $g$ is the bare coupling constant, we have
\beq
H_{\mathrm{Q}}[\psi, A; \beta] = H_{\mathrm{Q}}[\widetilde{\psi}, A; \beta] + \ii \mu \int_{\beta} \dd^3 x \, \widetilde{\psi}^{\dagger}(\vx) \alpha_3 \widetilde{\psi}(\vx) \label{Gl: DiracHamiltonian1}
\eeq
with\footnote{For finite baryon density the chemical potential would also receive a real part.}
\beq
\label{701-7}
\mu = \ii \frac{\pi - \varphi}{\beta}  \, .
\eeq
From the structure of the Dirac Hamiltonian (\ref{Gl: DiracHamiltonian}), it is clear that in the present Hamiltonian approach to finite temperatures a (real) chemical potential adds an imaginary part to the spatial momentum parallel to the compactified dimension. Thus, a pure imaginary chemical potential (\ref{701-7}) shifts this momentum by a real contribution
\beq
\label{707-8}
p_{\nf} = \Omega_{\nf} + \ii \mu = \omega_{\nf} + \frac{\varphi}{\beta} \, ,
\eeq
where $\Omega_{\nf}$ and $\omega_{\nf}$ are the fermionic and bosonic Matsubara frequencies defined in eq.~(\ref{287-f2}). The Fourier decomposition
of a function $f$ fulfilling the boundary condition (\ref{680-3}) is therefore given by
\beq
f(\vx) = \int_{\beta} \da^3 p \, \exp\bigl(\ii (\vp_{\perp} + p_{\nf} \he_3) \cdot \vx\bigr) f(\vp_{\perp}, p_{\nf})
\eeq
where the integration measure $\int_{\beta} \da^3 p$ was introduced in eq.~(\ref{305-f2a-x1}). After Poisson resummation (\ref{Gl: PoissonSummenformel}) we find
\beq
\label{382-f3a}
f(\vx) = \int \da^3 p \, \exp(\ii \vp \cdot \vx) f(\vp) \sum_{l = - \infty}^{\infty} \exp\left(\ii l \beta \left[p_3 - \frac{\varphi}{\beta}\right]\right) .
\eeq
This relation will be frequently used below.

\section{QCD in Coulomb gauge} \label{sectIV}

The variational Hamiltonian approach to QCD developed in ref.~\cite{Feuchter2004, *Feuchter2004a, *Feuchter2005} for pure Yang--Mills theory and extended in refs.~\cite{Pak2013, *Pak2012a, QCDT0, QCDT0Rev} to the quark sector can be equally well formulated on the spatial manifold $\mathbb{R}^2 \times S^1(\beta)$ in order to treat QCD at finite temperature in the way described in section \ref{secII}. For the Yang--Mills sector, this was already done in ref.~\cite{Heffner2015}. Therefore, below we will focus mainly on the quark sector.

\subsection{The gauge fixed Hamiltonian}\label{subsecA}

On $\mathbb{R}^2 \times S^1(\beta)$, the QCD Hamiltonian in Coulomb gauge $\nabla \cdot \vA = \nabla_\perp \cdot \vA_\perp + \partial_3 A_3 = 0$ reads
\beq
\label{735-1}
H = H_{\mathrm{YM}} + H_{\mathrm{Q}} + H_{\mathrm{C}} \, ,
\eeq
where 
\beq
\label{740-2}
H_{\mathrm{YM}} = \frac{1}{2} \int_{\beta} \dd^3 x \, \Bigl(J^{- 1}[A] \vPi(\vx) J[A] \vPi(\vx) + \vB^2(\vx)\Bigr)
\eeq
is the Hamiltonian of the transversal gluon fields. Here,
\beq
\label{416-4}
B_k^a(\vx) = \varepsilon_{k l m} \left(\partial_l A^a_m(\vx) - \frac{g}{2} f^{a b c} A_l^b(\vx) A_m^c(\vx)\right)
\eeq
is the non-Abelian chromomagnetic field with $f^{a b c}$ being the structure constant of the color group. Furthermore,
\beq
\Pi_k^a(\vx) = \frac{\delta}{\ii \delta A_k^a(\vx)}
\eeq
denotes the momentum operator, which represents the chromoelectric field, and
\beq
J[A] = \det\bigl({\hat{G}}^{-1}\bigr) \label{Gl: FaddeevPopov}
\eeq
is the Faddeev--Popov determinant in Coulomb gauge. The Faddeev--Popov operator
\beq
\bigl(\hat{G}^{-1}\bigr)^{a b}(\vx, \vy) := \bigl(-\nabla \cdot \hat{\vD}\bigr)^{a b}(\vx, \vy)
\eeq
contains the covariant derivative in the adjoint representation
\beq
\hat{D}^{a b}_k(\vx) = \delta^{ab} \partial_k - g f^{acb} A^c_k(\vx) \, . \label{Gl: KovAbleitung}
\eeq
The second term on the r.h.s.~of eq.~(\ref{735-1}), $H_{\mathrm{Q}}$, is the Hamiltonian (\ref{Gl: DiracHamiltonian}) of the quark field $\psi$ interacting with the transversal gluon field $\vA$. Finally,
\beq
H_{\mathrm{C}} = \frac{g^2}{2} \int_{\beta} \dd^3 x \int_{\beta} \dd^3 y \, J^{- 1}[A] \rho^a(\vx) J[A] \hat{F}^{ab}(\vx, \vy) 
\rho^b(\vy) \, , \label{Gl: Coulombterm}
\eeq
is the so-called Coulomb term which arises from the longitudinal part of the kinetic energy of the gluons after the resolution of Gau\ss{}'s law. Here,
\beq
\hat{F}^{a b}(\vx, \vy) = \int_{\beta} \dd^3 z \, \hat{G}^{a c}(\vx, \vz) (-\Delta_{\vz}) \hat{G}^{c b}(\vz, \vy) \label{Gl: Coulombkern}
\eeq
is the Coulomb kernel. Furthermore, $\rho = \rho_{\mathrm{Q}} + \rho_{\mathrm{YM}}$ is the total color charge which contains both the charge of the quarks
\beq
\rho^a_{\mathrm{Q}}(\vx) = \psi^\dagger(\vx) t_a \psi(\vx) \label{Gl: FarbdichteQ}
\eeq
and also the color charge of the gluons
\beq
\rho^a_{\mathrm{YM}}(\vx) = f^{abc} \vA^b(\vx) \cdot \vPi^c(\vx) \, . \label{Gl: FarbdichteYM}
\eeq
Since we are mainly interested in the quark sector, we will replace the Coulomb kernel $\hat{F}$ (\ref{Gl: Coulombkern}) by its Yang--Mills vacuum expectation value
\beq
g^2 \langle \hat{F}^{a b}(\vx, \vy) \rangle_{\mathrm{YM}} = \delta^{a b} V_{\mathrm{C}}(|\vx - \vy|) \label{Gl: Coulombkern3}
\eeq
which represents the static quark potential $V_{\mathrm{C}}$. This is correct when one restricts oneself up to including two-loop terms in the vacuum expectation value of the Hamiltonian, as we will do here, see below. The variational calculation in the Yang--Mills sector \cite{Feuchter2004, *Feuchter2004a, *Feuchter2005} shows that this potential can be nicely fitted by a linear combination of a linear potential plus an ordinary Coulomb potential
\beq
\label{410-f8-1}
V_{\mathrm{C}}(|\vx - \vy|) = -\sigma_{\mathrm{C}} \vert \vec{x} - \vec{y} \vert + \frac{\alpha_{\mathrm{S}}}{\vert \vec{x} - \vec{y} \vert} = V_{\mathrm{C}}^{\mathrm{IR}}(|\vx - \vy|) + V_{\mathrm{C}}^{\mathrm{UV}}(|\vx - \vy|) \, , \quad \quad \alpha_{\mathrm{S}} = \frac{g^2}{4 \pi}
\eeq
where $\sigma_{\mathrm{C}}$ is the Coulomb string tension. This quantity has been measured on the lattice and one finds values in the range
\beq
\label{915-32-1}
\sigma_{\mathrm{C}} / \sigma = 2 \ldots 4
\eeq
where $\sigma = (440 \, \mathrm{MeV})^2$ is the ordinary Wilson string tension \cite{Burgio2012, GOZ2004, Voigt2008}.

\subsection{The variational approach}\label{subsecB}

In ref.~\cite{Feuchter2004, *Feuchter2004a, *Feuchter2005}, the pure Yang--Mills theory was treated within a variational approach at zero temperature (on the spatial manifold $\mathbb{R}^3$) using a Gaussian type of ansatz for the vacuum wave functional. In ref.~\cite{Heffner2015}, this approach was extended to finite temperature by compactifying a spatial dimension, see section \ref{secII}. For the Yang--Mills vacuum wave functional on $\mathbb{R}^2 \times S^1(\beta)$, the following ansatz was used
\beq
\label{413-f4}
\phi_{\mathrm{YM}}[A] = J^{-\frac{1}{2}}[A] \exp\left[-\frac{1}{2} \int_{\beta} \dd^3 x \int_{\beta} \dd^3 y \, A_k^a(\vx) \hat{\omega}_{k l}^{a b} (\vx, \vy) A_l^b(\vy) \right] \equiv J^{-\frac{1}{2}}[A] \widetilde{\phi}_{\mathrm{YM}}[A]
\eeq
where
\beq
\label{413-f4-1}
\omega_{k l} = t^\perp_{k l} \omega_\perp + t^{||}_{k l} \omega_{||} 
\eeq
contains two independent variational kernels, $\omega_\perp$ and $\omega_{||}$, corresponding to the transversal projector in the subspace $\mathbb{R}^2$ orthogonal to the compactified dimension,
\beq
\label{425-f4-2}
t^\perp_{k l} = (1 - \delta_{k 3}) \left(\delta_{k l} - \frac{\partial_k \partial_l}{\Delta_\perp}\right) (1 - \delta_{l 3} ) \, ,
\eeq
and to its orthogonal complement
\beq
\label{431-f4-3}
t^{||}_{k l} = t_{k l} - t^\perp_{k l} \, .
\eeq
Here $t_{k l} = \delta_{k l} - \frac{\partial_k \partial_l}{\Delta}$ is the usual transversal projector in $\mathbb{R}^3$. With the ansatz (\ref{413-f4}), one finds for the gluon propagator of pure Yang--Mills theory
\beq
\label{440-f4-2-1}
\langle A^a_k(\vx) A^b_l(\vy) \rangle_{\mathrm{YM}} = \frac{1}{2} \delta^{a b} \left(t^\perp_{k l} \omega^{- 1}_\perp + t^{||}_{k l} \omega^{-1}_{||}\right)(\vx, \vy) \, .
\eeq
Minimization of the Yang--Mills energy density (which was calculated up to two loops) with respect to the two variational kernels $\omega_{\perp, ||}$ yields two coupled (gap) equations which were solved numerically in ref.~\cite{Heffner2015}. The obtained solutions show a finite temperature deconfinement phase transition which manifests itself in a decrease of the infrared exponents of the ghost form factor and the gluon energy. Furthermore, for $T \to 0$ the two kernels $\omega_\perp$ and $\omega_{||}$ both approach the $T = 0$ solution $\omega$ found in ref.~\cite{Feuchter2004, *Feuchter2004a, *Feuchter2005}. This zero temperature solution yields a gluon propagator (\ref{440-f4-2-1}) which is in good agreement with the lattice data and can be fitted by the Gribov formula \cite{Gribov1978}
\beq
\omega(p) = \sqrt{p^2 + \frac{m^4}{p^2}} \label{Gl: Gribov}
\eeq
with the Gribov mass $m \approx 880 \, \mathrm{MeV} \approx 2 \sqrt{\sigma}$ \cite{BQR2009}. The finite temperature solutions $\omega_{\perp}$, $\omega_{||}$ obtained in ref.~\cite{Heffner2015} differ from the zero temperature solution only in the small momentum regime. Since we are mainly interested here in the quark sector, we will use the zero temperature Yang--Mills solution parametrized by the Gribov formula (\ref{Gl: Gribov}) as input for the quark sector. This will simplify the numerical calculations considerably.

We will also ignore here the part of the Coulomb Hamiltonian $H_{\mathrm{C}}$ (\ref{Gl: Coulombterm}) describing the coupling of the color charge density of the quarks $\rho_{\mathrm{Q}}$ (\ref{Gl: FarbdichteQ}) with that of the gluons $\rho_{\mathrm{YM}}$ (\ref{Gl: FarbdichteYM}). This term contributes only in higher than second order in the number of loops. The part of the QCD Hamiltonian (\ref{735-1}) which contains the quark field is then given by
\beq
\label{802-10}
\bar{H}_{\mathrm{Q}} = H_{\mathrm{Q}} + H_{\mathrm{C}}^{\mathrm{Q}}
\eeq
where $H_{\mathrm{Q}}$ is defined by eq.~(\ref{Gl: DiracHamiltonian}) and $H_{\mathrm{C}}^{\mathrm{Q}}$ is obtained from eq.~(\ref{Gl: Coulombterm}) 
by replacing the total color charge density $\rho$ by that of the quarks, $\rho_{\mathrm{Q}}$ (\ref{Gl: FarbdichteQ}).

In refs.~\cite{Pak2013, *Pak2012a, QCDT0, QCDT0Rev}, the variational approach of ref.~\cite{Feuchter2004, *Feuchter2004a, *Feuchter2005} was extended to the quark sector. We follow here refs.~\cite{QCDT0, QCDT0Rev} and use the following variational ansatz for the quark part of the vacuum wave functional\footnote{Throughout this paper we use the bracket notation in the quark sector but the coordinate representation in the Yang--Mills sector.}
\beq
\label{812-11}
\vert \phi_{\mathrm{Q}}[A] \rangle = \exp\left[-\int_{\beta} \dd^3 x \int_{\beta} \dd^3 y \, \psi_+^{\dagger}(\vx) K(\vx, \vy) \psi_-(\vy)\right] \vert 0 \rangle
\eeq
where
\beq
\label{5124-41}
K(\vx, \vy) = \gamma_0 S(\vx - \vy) + g \int_{\beta} \dd^3 z \, \Bigl(V(\vx, \vy; \vz) + \gamma_0 W(\vx, \vy; \vz)\Bigr) \valpha \cdot \vA(\vz) \, .
\eeq
Here $\psi_{\pm}$ denotes the positive and negative energy components of the quark field
\beq
\label{500-f5}
\psi(\vx) =  \psi_+(\vx) + \psi_-(\vx)  \, ,
\eeq
see appendix \ref{Anh: Impulsdarstellung}. Furthermore, $\vert 0 \rangle$ is the bare Dirac vacuum satisfying 
\beq
\label{506-f5-1}
\psi_+(\vx) \vert 0 \rangle = 0 = \psi_-^{\dagger}(\vx) \vert 0 \rangle \, .
\eeq
Finally, $S$, $V$ and $W$ are variational kernels. We have suppressed here all indices of the quark field, in particular, the flavor index. In principle, the variational kernels are flavor dependent when the different electroweak quark masses of the (light) quark flavors are included. In the present paper, we consider only chiral quarks so that the flavor dependence can be ignored. Obviously, our quark wave functional $\vert \phi_{\mathrm{Q}}[A] \rangle$ (\ref{812-11}) reduces to the bare Dirac vacuum $\vert 0 \rangle$ for $K = 0$, i.e.~for $S = V = W = 0$. The ansatz (\ref{812-11}) is an extension of previous treatments of the  quark sector in Coulomb gauge. For $W = 0$ it reduces to the ansatz considered in ref.~\cite{Pak2013, *Pak2012a}, while for $V = W = 0$ it becomes the BCS type of wave functional used in ref.~\cite{Adler1984}. In ref.~\cite{QCDT0Rev}, it was shown that the inclusion of the explicit coupling of the quarks to the transversal gluons in the vacuum wave functional increases considerably the amount of spontaneous breaking of chiral symmetry (compared to the BCS case of $V = W = 0$). By including also the coupling term $\sim W$ in eq.~(\ref{5124-41}), the resulting gap equation for the scalar kernel $S$ becomes free of UV divergences: As we will see shortly, the linear UV divergences arising from the $W$-term exactly cancel those originating from the $V$-term. Furthermore, the logarithmic UV divergences induced by both the $V$- and $W$-vertex cancel the UV divergence arising from the Coulomb term $H_{\mathrm{C}}^{\mathrm{Q}}$ (\ref{802-10}).

For later application we also quote the norm of the quark wave functional (\ref{812-11}):
\beq
I[A] = \langle \phi_{\mathrm{Q}}[A] \vert \phi_{\mathrm{Q}}[A] \rangle = \det(\Id + K^{\dagger} K) \, . \label{Gl: Fermideterminante}
\eeq
We will be later interested in the variational solution of the quark sector for quark fields satisfying the phase dependent boundary conditions (\ref{680-3}) for the compactified spatial dimension in order to calculate the dual condensate (\ref{Gl: DualesKondensat}). For such quark fields, continuity of the wave functional at $x_3 = \beta$ requires that the quark kernel $K(\vx, \vy)$ also satisfies phase dependent boundary conditions:
\begin{subequations}
\begin{align}
K (\vx + \beta \he_3, \vy) &= \mathrm{e}^{\ii \varphi} K (\vx, \vy) \, , \\
K (\vx, \vy + \beta \he_3) &= \mathrm{e}^{-\ii \varphi} K (\vx, \vy) \, .
\end{align}
\end{subequations}
For the total ground state wave functional of QCD, we use an ansatz analogous to the one in refs.~\cite{QCDT0, QCDT0Rev}
\beq
\label{538-41}
\vert \phi[A] \rangle = \Nc I^{-\frac{1}{2}}[A] \phi_{\mathrm{YM}}[A] \vert \phi_{\mathrm{Q}}[A] \rangle = \Nc I^{-\frac{1}{2}}[A] J^{-\frac{1}{2}}[A] \widetilde{\phi}_{\mathrm{YM}}[A] \vert \phi_{\mathrm{Q}}[A] \rangle
\eeq
where $\phi_{\mathrm{YM}}$ (\ref{413-f4}) is the wave functional of the Yang--Mills sector, $I$ is the fermion determinant (\ref{Gl: Fermideterminante}) and $\Nc$ denotes a normalization factor. Note that with our ansatz the ghost and quark determinants enter the total QCD vacuum wave functional (\ref{538-41}) on an equal footing.

\subsection{Variational equations of the quark sector}\label{subsecC}

\begin{figure}%
\centering%
\includegraphics[width=0.15\linewidth]{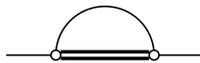}%
\caption{Diagrammatic representation of the contribution (\ref{700-56}) of the Coulomb interaction to the quark gap equation (\ref{Gl: Gapgleichungresummiert}). Straight lines stand for quark propagators while the Coulomb kernel $V_{\mathrm{C}}$ (\ref{Gl: Coulombkern3}) is illustrated by a double line.}%
\label{Abb: FeynmanCoulombGapgl}%
\end{figure}

With the ansatz (\ref{538-41}) for the QCD vacuum wave functional we calculate the expectation value of $H$ (\ref{735-1}) up to two loops\footnote{The loop counting is here as usual except that the loops are formed with the dressed instead of bare propagators.} and carry out the variation with respect to the kernels $S$, $V$ and $W$ of the quark wave functional. For the gluon propagator we will thereby assume the Gribov formula (\ref{Gl: Gribov}) which nicely fits the zero temperature lattice data. Furthermore, for simplicity we will consider only chiral quarks ($m_{\mathrm{Q}} = 0$). In the evaluation of the ground state energy it is convenient to take the fermionic expectation value first. This yields
\beq
\langle H \rangle = |\Nc|^2 \int \Dc A \, \widetilde{\phi}_{\mathrm{YM}}^*[A] \langle \widetilde{H} \rangle_{\mathrm{Q}} \, \widetilde{\phi}_{\mathrm{YM}}[A] \label{Gl: EnergieEW}
\eeq
where
\beq
\label{580-6}
\langle O \rangle_{\mathrm{Q}} = I^{-1}[A] \langle \phi_{\mathrm{Q}} \vert O \vert \phi_{\mathrm{Q}} \rangle
\eeq
is the (normalized) fermionic expectation value in the state (\ref{812-11}) and we have defined\footnote{Note that the Faddeev--Popov determinant $J$ [Eq.~(\ref{Gl: FaddeevPopov})] and the quark determinant $I$ [Eq.~(\ref{Gl: Fermideterminante})] are both functionals of the gauge field and hence do not commute with the momentum operator and, thus, not with $H$. Let us also mention that the scalar product in the Hilbert space of the gauge field contains the Faddeev--Popov determinant in the integration measure after fixing to Coulomb gauge. However, due to our ansatz (\ref{413-f4}) for the Yang--Mills vacuum wave functional the Faddeev--Popov determinant drops out from the integration measure in eq.~(\ref{Gl: EnergieEW}).}
\beq
\widetilde{H} = J^{\frac{1}{2}}[A] I^{\frac{1}{2}}[A] \, H \, J^{-\frac{1}{2}}[A] I^{-\frac{1}{2}}[A] \, .
\eeq
These calculations can be done completely analogously to the zero temperature case \cite{QCDT0} and are conveniently carried out in momentum space, where the momenta along the compactified spatial dimension are given by the corresponding Matsubara frequencies (\ref{707-8}), see appendix \ref{Anh: Impulsdarstellung} for the momentum space representation of the wave functional. From the minimization of the ground state energy (\ref{Gl: EnergieEW}), one finds the equations of motion in terms of the Matsubara frequencies, see appendix \ref{Anh: Gapgleichung}. For convenience, the occurring Matsubara sums are Poisson resummed using eq.~(\ref{Gl: PoissonSummenformel}). Then, after shifting both the external and internal momenta by $-\varphi / \beta \he_3$, the quark gap equation for the variational kernel $S(\vp)$ reads
\beq
p S(\vp) = I_{\mathrm{C}}(\vp) + I_{V V}(\vp) + I_{W W}(\vp) + I_{V \mathrm{Q}}(\vp) + I_{W \mathrm{Q}}(\vp) + I_{V E}(\vp) + I_{W E}(\vp) \label{Gl: Gapgleichungresummiert}
\eeq
where the various terms on the r.h.s.~are all one-loop integrals. The first loop term,
\beq
\label{700-56}
I_{\mathrm{C}}(\vp) = \frac{C_{\mathrm{F}}}{2} \int \da^3 q \, \sum_{l = -\infty}^{\infty} \exp\left(\ii l \beta \left[q_3 - \frac{\varphi}{\beta}\right]\right) V_{\mathrm{C}}(|\vq - \vp|) P(\vq) \left[S(\vq) \Bigl(1 - S^2(\vp)\Bigr) - S(\vp) \Bigl(1 - S^2(\vq)\Bigr) \hp \cdot \hq\right]
\eeq
($\hp \equiv \vp / p$), arises exclusively from the Coulomb interaction $H_{\mathrm{C}}^{\mathrm{Q}}$ (\ref{802-10}). Here $C_{\mathrm{F}} = (\Ncc^2 - 1) / 2 \Ncc$ is the quadratic Casimir and we have defined
\beq
\label{606-f7b}
P (\vp) = \frac{1}{1 + S^2 (\vp)} \, .
\eeq
As in the zero temperature case \cite{QCDT0}, all loop integrals on the r.h.s.~of eq.~(\ref{Gl: Gapgleichungresummiert}) allow for an interpretation in terms of Feynman diagrams. For the contribution of the Coulomb interaction (\ref{700-56}), this is given in fig.~\ref{Abb: FeynmanCoulombGapgl}.

The next two terms in the gap equation (\ref{Gl: Gapgleichungresummiert}) arise from the free single particle Dirac Hamiltonian $H_{\mathrm{Q}}^0$ (\ref{Gl: DiracHamiltonian}) and the two quark-gluon vertices $V$ and $W$ in the vacuum wave functional (\ref{812-11}) with (\ref{5124-41})
\begin{align}
I_{V V}(\vp) &= -\frac{C_{\mathrm{F}}}{2} g^2 \int \da^3 q \, \sum_{l = -\infty}^{\infty} \exp\left(\ii l \beta \left[q_3 - \frac{\varphi}{\beta}\right]\right) \frac{V^2(\vq, -\vp)}{\omega\bigl(\vert \vq - \vp\vert\bigr)} X\bigl(\vq, -\vp\bigr) P(\vq) \nonumber \\
&\phantom{=}\,\, \phantom{-\frac{C_{\mathrm{F}}}{2} g^2 \int \da^3 q \,} \times \Bigl[q P(\vq) \Bigl\{S(\vq) \Bigl(1 - S^2(\vp)\Bigr) - S(\vp) \Bigl(1 - S^2(\vq)\Bigr)\Bigr\} \nonumber \\
&\phantom{=}\,\, \phantom{-\frac{C_{\mathrm{F}}}{2} g^2 \int \da^3 q \, \times \Bigl[} + p P(\vp) \Bigl\{S(\vq) \Bigl(1 - 3 S^2(\vp)\Bigr) - S(\vp) \Bigl(3 - S^2(\vp)\Bigr)\Bigr\}\Bigr] \, , \label{Gl: FrDiracGapglV} \\
I_{W W}(\vp) &= -\frac{C_{\mathrm{F}}}{2} g^2 \int \da^3 q \, \sum_{l = -\infty}^{\infty} \exp\left(\ii l \beta \left[q_3 - \frac{\varphi}{\beta}\right]\right) \frac{W^2(\vq, -\vp)}{\omega\bigl(\vert \vq - \vp\vert\bigr)} Y\bigl(\vq, -\vp\bigr) P(\vq) \nonumber \\
&\phantom{=}\,\, \phantom{-\frac{C_{\mathrm{F}}}{2} g^2 \int_{\beta} \da^3 q \,} \times \Bigl[q P(\vq) \Bigl\{S(\vq) \Bigl(-1 + S^2(\vp)\Bigr) - S(\vp) \Bigl(1 - S^2(\vq)\Bigr)\Bigr\} \nonumber \\
&\phantom{=}\,\, \phantom{-\frac{C_{\mathrm{F}}}{2} g^2 \int_{\beta} \da^3 q \, \times \Bigl[} + p P(\vp) \Bigl\{S(\vq) \Bigl(-1 + 3 S^2(\vp)\Bigr) - S(\vp) \Bigl(3 - S^2(\vp)\Bigr)\Bigr\}\Bigr] \label{Gl: FrDiracGapglW}
\end{align}
\begin{figure}%
\centering%
\parbox{0.125\linewidth}{%
\centering%
\includegraphics[width=\linewidth]{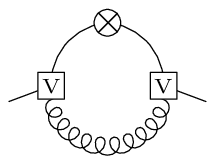} \\%
(a) %
}%
\hspace{0.1\linewidth}%
\parbox{0.125\linewidth}{%
\centering%
\includegraphics[width=\linewidth]{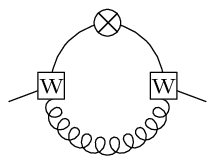} \\%
(b) %
}%
\caption{Diagrammatic representation of the one-loop contributions of the free Dirac Hamiltonian to the quark gap equation, (a) eq.~(\ref{Gl: FrDiracGapglV}) and (b) eq.~(\ref{Gl: FrDiracGapglW}). A curly line stands for the static gluon propagator, while the free Dirac operator and the vector kernels $V$, $W$ are marked by a crossed circle and labeled squares, respectively.}%
\label{Abb: FeynmanFrDiracGapgl}%
\end{figure}%
and are diagrammatically illustrated in fig.~\ref{Abb: FeynmanFrDiracGapgl}. Here we have introduced the abbreviations
\begin{subequations}%
\label{Gl: Winkelfaktoren} %
\begin{align}%
X(\vp, \vq) &= 1 - \Bigl[\hp \cdot \hpq\Bigr] \Bigl[\hq \cdot \hpq\Bigr] \\
Y(\vp, \vq) &= 1 + \Bigl[\hp \cdot \hpq\Bigr] \Bigl[\hq \cdot \hpq\Bigr] \, .
\end{align}%
\end{subequations}%
The quark-gluon coupling in the Dirac-Hamiltonian $H_{\mathrm{Q}}^A$ (\ref{Gl: DiracHamiltonian}) gives rise to the following loop terms
\begin{align}
I_{V \mathrm{Q}}(\vp) &= \frac{C_{\mathrm{F}}}{2} g^2 \int \da^3 q \, \sum_{l = -\infty}^{\infty} \exp\left(\ii l \beta \left[q_3 - \frac{\varphi}{\beta}\right]\right) \frac{V(\vq, -\vp)}{\omega\bigl(\vert \vq - \vp\vert\bigr)} X\bigl(\vq, -\vp\bigr) P(\vq) \Bigl[S(\vq) - 2 S(\vp) - S(\vq) S^2(\vp)\Bigr] \, , \label{Gl: KopplungstermGapglV} \\
I_{W \mathrm{Q}}(\vp) &= \frac{C_{\mathrm{F}}}{2} g^2 \int \da^3 q \, \sum_{l = -\infty}^{\infty} \exp\left(\ii l \beta \left[q_3 - \frac{\varphi}{\beta}\right]\right) \frac{W(\vq, -\vp)}{\omega\bigl(\vert \vq - \vp\vert\bigr)} Y\bigl(\vq, -\vp\bigr) P(\vq) \Bigl[1 - 2 S(\vq) S(\vp) - S^2(\vp)\Bigr] \label{Gl: KopplungstermGapglW}
\end{align}
\begin{figure}%
\centering%
\parbox{0.125\linewidth}{%
\centering%
\includegraphics[width=\linewidth]{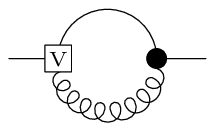} \\%
(a) %
}%
\hspace{0.1\linewidth}%
\parbox{0.125\linewidth}{%
\centering%
\includegraphics[width=\linewidth]{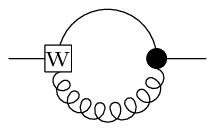} \\%
(b) %
}%
\caption{Diagrammatic representation of the contributions of the quark-gluon coupling in the Dirac Hamiltonian to the quark gap equation, (a) eq.~(\ref{Gl: KopplungstermGapglV}) and (b) eq.~(\ref{Gl: KopplungstermGapglW}). The filled dot stands for the bare quark-gluon vertex in the Dirac Hamiltonian $H_{\mathrm{Q}}^A$ (\ref{Gl: DiracHamiltonian}).}%
\label{Abb: FeynmanKopplungstermGapgl}%
\end{figure}%
which are illustrated in fig.~\ref{Abb: FeynmanKopplungstermGapgl}. Finally, there are quark loops arising from the action of the kinetic energy operator of the gluons (\ref{740-2}) on the quark wave functional (\ref{812-11})
\begin{align}
I_{V E}(\vp) &= \frac{C_{\mathrm{F}}}{2} g^2 S(\vp) \int \da^3 q \, \sum_{l = -\infty}^{\infty} \exp\left(\ii l \beta \left[q_3 - \frac{\varphi}{\beta}\right]\right) V^2(\vq, -\vp) X\bigl(\vq, -\vp\bigr) P(\vq) \, , \label{Gl: KinetischeEnergieGapglV} \\
I_{W E}(\vp) &= \frac{C_{\mathrm{F}}}{2} g^2 S(\vp) \int \da^3 q \, \sum_{l = -\infty}^{\infty} \exp\left(\ii l \beta \left[q_3 - \frac{\varphi}{\beta}\right]\right) W^2(\vq, -\vp) Y\bigl(\vq, -\vp\bigr) P(\vq) \label{Gl: KinetischeEnergieGapglW}
\end{align}
\begin{figure}%
\centering%
\parbox{0.125\linewidth}{%
\centering%
\includegraphics[width=\linewidth]{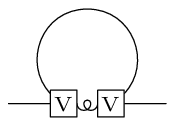} \\%
(a) %
}%
\hspace{0.1\linewidth}%
\parbox{0.125\linewidth}{%
\centering%
\includegraphics[width=\linewidth]{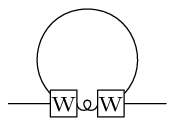} \\%
(b) %
}%
\caption{Diagrammatic representation of the contributions of the kinetic energy of the transversal gluons to the quark gap equation, (a) eq.~(\ref{Gl: KinetischeEnergieGapglV}) and (b) eq.~(\ref{Gl: KinetischeEnergieGapglW}). The short wavy line represents the direct connection of two gluonic legs of the quark-gluon {``vertices''} $V$, $W$.}%
\label{Abb: FeynmanKinEnGapgl}%
\end{figure}%
which are diagrammatically illustrated in fig.~\ref{Abb: FeynmanKinEnGapgl}.

Note that, due to the different Dirac structure of the $V$- and $W$-vertex in the vacuum wave functional (\ref{5124-41}), there is no one-loop term with both one $V$- and one $W$-vertex entering the gap equation (\ref{Gl: Gapgleichungresummiert}). Furthermore, in all the above given one-loop terms the $l = 0$ contribution yields the corresponding zero temperature expressions \cite{QCDT0, QCDT0Rev}, which contain all UV singular pieces, while the temperature dependent contributions $(l \neq 0)$ are UV finite.

The variational equations for the quark-gluon coupling kernels $V$ and $W$ can be solved explicitly in terms of the scalar kernel $S$. One finds
\begin{align}
V(\vp, -\vq) &= \frac{1 + S(\vp) S(\vq)}{p P(\vp) \Bigl(1 - S^2(\vp) + 2 S(\vp) S(\vq)\Bigr) + q P(\vq) \Bigl(1 - S^2(\vq) + 2 S(\vp) S(\vq)\Bigr) + \omega(|\vp - \vq|)} \, , \label{Gl: VKern1} \\
W(\vp, -\vq) &= \frac{S(\vp) + S(\vq)}{p P(\vp) \Bigl(1 - S^2(\vp) - 2 S(\vp) S(\vq)\Bigr) + q P(\vq) \Bigl(1 - S^2(\vq) - 2 S(\vp) S(\vq)\Bigr) + \omega(|\vp - \vq|)} \, . \label{Gl: WKern1}
\end{align}
Note that, contrary to the one-loop terms in the gap equation (\ref{Gl: Gapgleichungresummiert}) for the scalar kernel $S$, these expressions do not contain the sum over the oscillating phases resulting from the Poisson resummation, see e.g.~eq.~(\ref{700-56}). The reason is that the kernels $V$, $W$ depend on two momentum variables and by taking the variation of the energy (which contains at most two-loop terms) with respect to these kernels both loop momenta are fixed to the external momenta.

Inserting the explicit expressions for $V$ (\ref{Gl: VKern1}) and $W$ (\ref{Gl: WKern1}) into the gap equation (\ref{Gl: Gapgleichungresummiert}) and assuming the scalar kernel $S$ to vanish rapidly in the UV (which should be the case due to asymptotic freedom), one finds that the loop terms in the gap equation (\ref{Gl: Gapgleichungresummiert}) arising from the $V$-vertex have the following (divergent) UV behavior \cite{QCDT0, QCDT0Rev}
\beq
I_{V V}^{\mathrm{div}}(\vp) + I_{V \mathrm{Q}}^{\mathrm{div}}(\vp) + I_{V E}^{\mathrm{div}}(\vp) = \frac{C_{\mathrm{F}}}{16 \pi^2} g^2 S(\vp) \left[-2 \Lambda + p \ln \frac{\Lambda}{\mu} \left(-\frac{2}{3} + 4 P(\vp)\right)\right] \label{Gl: UVLimesV}
\eeq
where $\Lambda$ is the UV cutoff and $\mu$ is an arbitrary momentum scale.\footnote{We follow here the standard nomenclature and denote the momentum scale as well as the chemical potential by the Greek letter $\mu$. This should, however, not give rise to any confusion.} For the loops arising from the $W$-vertex, one finds instead the UV behavior
\beq
I_{W W}^{\mathrm{div}}(\vp) + I_{W \mathrm{Q}}^{\mathrm{div}}(\vp) + I_{W E}^{\mathrm{div}}(\vp) = \frac{C_{\mathrm{F}}}{16 \pi^2} g^2 S(\vp) \left[2 \Lambda + p \ln \frac{\Lambda}{\mu} \left(\frac{10}{3} - 4 P(\vp)\right)\right] . \label{Gl: UVLimesW}
\eeq
Under the same assumptions, one finds that both vector kernels $V$ (\ref{Gl: VKern1}) and $W$ (\ref{Gl: WKern1}) have the same UV behavior for fixed $q$ and large $p$. Comparing eqs.~(\ref{Gl: UVLimesV}) and (\ref{Gl: UVLimesW}) for the divergences arising from the $V$- and $W$-contributions, one observes that the linear divergent terms come with opposite sign and, as a result, the linear UV divergences cancel in the gap equation (\ref{Gl: Gapgleichungresummiert}) for the scalar kernel $S$. Furthermore, the Coulomb term $I_{\mathrm{C}}$ (\ref{700-56}) has the UV behavior
\beq
I_{\mathrm{C}}^{\mathrm{div}}(\vp) = -\frac{C_{\mathrm{F}}}{6 \pi^2} g^2 p S(\vp) \ln\frac{\Lambda}{\mu}
\eeq
which exactly cancels the logarithmic UV divergences of $I_{V V}^{\mathrm{div}}$ and $I_{W W}^{\mathrm{div}}$. Thus, the gap equation (\ref{Gl: Gapgleichungresummiert}) is free of UV divergences. This is certainly a very pleasant feature of our ansatz (\ref{812-11}), (\ref{5124-41}) for the quark wave functional.

Finally, defining the effective quark mass\footnote{From the explicit form of the quark propagator in our approach at zero temperature, it is seen that $M(\vp)$ and $E(\vp)$ have the meaning of an effective mass and the quasi-particle energy, respectively, see ref.~\cite{QCDT0} and eq.~(\ref{Gl: StatProp}) below.}
\beq
M(\vp) = \frac{2 p S(\vp)}{1 - S^2(\vp)} \label{Gl: Massenfkt}
\eeq
and the quasi-particle energy
\beq
E(\vp) = \sqrt{p^2 + M^2(\vp)} \, , \label{Gl: Quasiteilchenenergie}
\eeq
the quark gap equation (\ref{Gl: Gapgleichungresummiert}) can be rewritten in the alternative form
\beq
M(\vp) = \Ic_{\mathrm{C}}(\vp) + \Ic_{V V}(\vp) + \Ic_{W W}(\vp) + \Ic_{V \mathrm{Q}}(\vp) + \Ic_{W \mathrm{Q}}(\vp) + \Ic_{V E}(\vp) + \Ic_{W E}(\vp) \label{Gl: GapgleichungMassenfunktion}
\eeq
which is more convenient for the numerical calculation. Here, the loop terms on the r.h.s~are given by the following expressions
\begin{align}
\Ic_{\mathrm{C}}(\vp) &= \frac{C_{\mathrm{F}}}{2} \int \da^3 q \, \sum_{l = -\infty}^{\infty} \exp\left(\ii l \beta \left[q_3 - \frac{\varphi}{\beta}\right]\right) V_{\mathrm{C}}(|\vq - \vp|) \frac{M(\vq) - M(\vp) \frac{\vq \cdot \vp}{p^2}}{E(\vq)} \, , \label{Gl: SchleifenintegralMC} \\
\Ic_{V V}(\vp) &= -\frac{C_{\mathrm{F}}}{2} g^2 \int \da^3 q \, \sum_{l = -\infty}^{\infty} \exp\left(\ii l \beta \left[q_3 - \frac{\varphi}{\beta}\right]\right) \frac{V^2(\vq, -\vp)}{\omega(|\vq - \vp|)} X(\vq, -\vp) \left\{-\frac{E(\vq) + q}{2 E(\vq)} M(\vp) \frac{E(\vp) + 2 p}{E(\vp)} \right. \nonumber \\
&\phantom{=}\,\, \phantom{-\frac{C_{\mathrm{F}}}{2} g^2 \int \da^3 q} \left. - q^2 \frac{E(\vq) + q}{2 E^2(\vq)} \frac{M(\vp)}{p} + \frac{M(\vq)}{2 E(\vq)} \frac{E(\vp) + p}{E(\vp)} \Bigl[-E(\vp) + 2 p\Bigr] + q M(\vq) \frac{E(\vq) + q}{2 E^2(\vq)}\right\} \, , \\
\Ic_{W W}(\vp) &= -\frac{C_{\mathrm{F}}}{2} g^2 \int \da^3 q \, \sum_{l = -\infty}^{\infty} \exp\left(\ii l \beta \left[q_3 - \frac{\varphi}{\beta}\right]\right) \frac{W^2(\vq, -\vp)}{\omega(|\vq - \vp|)} Y(\vq, -\vp) \left\{-\frac{E(\vq) + q}{2 E(\vq)} M(\vp) \frac{E(\vp) + 2 p}{E(\vp)} \right. \nonumber \\
&\phantom{=}\,\, \phantom{-\frac{C_{\mathrm{F}}}{2} g^2 \int \da^3 q} \left. - q^2 \frac{E(\vq) + q}{2 E^2(\vq)} \frac{M(\vp)}{p} - \frac{M(\vq)}{2 E(\vq)} \frac{E(\vp) + p}{E(\vp)} \Bigl[-E(\vp) + 2 p\Bigr] - q M(\vq) \frac{E(\vq) + q}{2 E^2(\vq)}\right\} \, , \\
\Ic_{V \mathrm{Q}}(\vp) &= \frac{C_{\mathrm{F}}}{2} g^2 \int \da^3 q \, \sum_{l = -\infty}^{\infty} \exp\left(\ii l \beta \left[q_3 - \frac{\varphi}{\beta}\right]\right) \frac{V(\vq, -\vp)}{\omega(|\vq - \vp|)} X(\vq, -\vp) \left[\frac{M(\vq)}{E(\vq)} - \frac{E(\vq) + q}{E(\vq)} \frac{M(\vp)}{p}\right] , \\
\Ic_{W \mathrm{Q}}(\vp) &= \frac{C_{\mathrm{F}}}{2} g^2 \int \da^3 q \, \sum_{l = -\infty}^{\infty} \exp\left(\ii l \beta \left[q_3 - \frac{\varphi}{\beta}\right]\right) \frac{W(\vq, -\vp)}{\omega(|\vq - \vp|)} Y(\vq, -\vp) \left[\frac{E(\vq) + q}{E(\vq)} - \frac{M(\vq)}{E(\vq)} \frac{M(\vp)}{p}\right] , \\
\Ic_{V E}(\vp) &= \frac{C_{\mathrm{F}}}{2} g^2 \frac{M(\vp)}{p} \int \da^3 q \, \sum_{l = -\infty}^{\infty} \exp\left(\ii l \beta \left[q_3 - \frac{\varphi}{\beta}\right]\right) V^2(\vq, -\vp) X(\vq, -\vp) \frac{E(\vq) + q}{2 E(\vq)} \, , \\
\Ic_{W E}(k) &= \frac{C_{\mathrm{F}}}{2} g^2 \frac{M(\vp)}{p} \int \da^3 q \, \sum_{l = -\infty}^{\infty} \exp\left(\ii l \beta \left[q_3 - \frac{\varphi}{\beta}\right]\right) W^2(\vq, -\vp) Y(\vq, -\vp) \frac{E(\vq) + q}{2 E(\vq)} \label{Gl: SchleifenintegralMWE}
\end{align}
where for the vector kernels
\beq
V(\vp, -\vq) = \frac{1 + \frac{E(\vp) - p}{M(\vp)} \frac{E(\vq) - q}{M(\vq)}}{\frac{p^2}{E(\vp)} \left[1 + \frac{M(\vp)}{p} \frac{E(\vq) - q}{M(\vq)}\right] + \frac{q^2}{E(\vq)} \left[1 + \frac{M(\vq)}{q} \frac{E(\vp) - p}{M(\vp)}\right] + \omega(|\vp - \vq|)}
\eeq
and
\beq
W(\vp, -\vq) = \frac{\frac{E(\vp) - p}{M(\vp)} + \frac{E(\vq) - q}{M(\vq)}}{\frac{p^2}{E(\vp)} \left[1 - \frac{M(\vp)}{p} \frac{E(\vq) - q}{M(\vq)}\right] + \frac{q^2}{E(\vq)} \left[1 - \frac{M(\vq)}{q} \frac{E(\vp) - p}{M(\vp)}\right] + \omega(|\vp - \vq|)} \, .
\eeq
holds.

The full temperature dependence as well as the dependence on the phase $\varphi$ of the boundary condition (\ref{680-3}) to the quark field is exclusively contained in the exponentials on the r.h.s.~of eq.~(\ref{Gl: GapgleichungMassenfunktion}). Furthermore, the $l = 0$  terms are obviously independent of the temperature $\beta^{- 1}$ and the phase $\varphi$, and yield precisely the zero temperature quark gap equation derived in ref.~\cite{QCDT0Rev}. For simplicity, in this paper in the numerical calculation of the order parameter to be given in section \ref{Abschn: Numerik}, we will use the zero temperature solution, i.e.~keep in eq.~(\ref{Gl: GapgleichungMassenfunktion}) only the $l = 0$ terms. This yields an effective quark mass (\ref{Gl: Massenfkt}) which depends on the modulus of the momentum $\vp$ only
\beq
\label{729-f9}
M(\vp) = M(p) \, .
\eeq
The influence of the remaining (temperature-dependent) terms $l \neq 0$ in eq.~(\ref{Gl: GapgleichungMassenfunktion}) will be subject to forthcoming investigations \cite*{[{}][{to be published}]ERV201X}.

\section{Chiral and dual quark condensate in the Hamiltonian approach in Coulomb gauge}\label{sectV}

Below, we calculate the chiral quark condensate and the dual quark condensate (dressed Polyakov loop) within the variational approach presented in the previous section.

\subsection{Quark condensate}

The quark condensate
\beq
\langle \bar{\psi}^m(\vx) \psi^m(\vx) \rangle_{\varphi} = -\tr\bigl(\gamma_0 G(\vx, \vx)\bigr) \label{Gl: ChiralesKondensat3}
\eeq
can be easily obtained once the static quark propagator 
\begin{align}
G_{i j}^{m n}(\vx, \vy) &= \frac{1}{2} \bigl\langle \Bigl[\psi_i^m(\vx), {\psi_j^n}^{\dagger}(\vy)\Bigr] \bigr\rangle = \delta^{m n} \int_{\beta} \da^3 p \, \exp\bigl(\ii (\vp_{\perp} + p_{\nf} \he_3) \cdot (\vx - \vy)\bigr) G_{i j}(\vp_{\perp}, p_{\nf}) \nonumber \\
&= \delta^{m n} \int \da^3 p \, \sum_{l = -\infty}^{\infty} \exp\left(\ii l \beta \left[p_3 - \frac{\varphi}{\beta}\right]\right) \exp\bigl(\ii \vp \cdot (\vx - \vy)\bigr) G_{i j}(\vp)
\end{align}
is known. To leading order, the quark propagator is given by \cite{QCDT0, QCDT0Rev}
\beq
G(\vp) = \frac{\valpha \cdot \vp + \beta M(\vp)}{2 E(\vp)} \label{Gl: StatProp}
\eeq
where $M(\vp)$ and $E(\vp)$ are defined by eqs.~(\ref{Gl: Massenfkt}) and (\ref{Gl: Quasiteilchenenergie}), respectively. With this expression, we find for the quark condensate (\ref{Gl: ChiralesKondensat3})
\beq
\langle \bar{\psi}^m(\vx) \psi^m(\vx) \rangle_\varphi = -2 \Ncc \int \da^3 p \, \sum_{l = -\infty}^{\infty} \exp\left(\ii l \beta \left[p_3 - \frac{\varphi}{\beta}\right]\right) \frac{M(\vp)}{E(\vp)} \, . \label{Gl: ChiralesKondensat}
\eeq

As mentioned at the end of section \ref{sectIV}, in the numerical calculation we will use the zero temperature solution of the quark gap equation which yields an $O(3)$ invariant quark mass function $M(\vp) = M(p)$. For such mass functions, the angular integrations in the quark condensate (\ref{Gl: ChiralesKondensat}) can be explicitly carried out\footnote{The series is not absolutely convergent, therefore it is a priori not clear if the order of integration and summation plays a role. However, our numerical calculations yield the same result, independent of the order in which the two operations are performed.} and we find
\begin{align}
\langle \bar{\psi}^m(\vx) \psi^m(\vx) \rangle_\varphi &= -\frac{\Ncc}{\pi^2} \int\limits_0^{\infty} \dd p \, \frac{p^2 M(p)}{E(p)} \left[1 + 2 \sum_{l = 1}^{\infty} \cos(l \varphi) \,\sinc(l \beta p) \right] \label{Gl: ChiralesKondensat1}
\end{align}
where
\beq
\label{906-121}
\sinc x \equiv \frac{\sin x}{x} \, .
\eeq
Obviously, the quark condensate $\langle \bar{\psi} \psi \rangle_\varphi$ (\ref{Gl: ChiralesKondensat1}) is a periodic function in the phase $\varphi$ with period $2 \pi$. From eq.~(\ref{Gl: ChiralesKondensat1}), one finds in the zero temperature limit of the quark condensate
\beq
\lim_{\beta \to \infty} \langle \bar{\psi}^m(\vx) \psi^m(\vx) \rangle_\varphi = -\frac{\Ncc}{\pi^2} \int\limits_0^{\infty} \dd p \, \frac{p^2 M(p)}{E(p)} \label{Gl: ChiralesKondensat0T} \, .
\eeq
This limit is independent of the phase $\varphi$ and agrees with the result obtained in ref.~\cite{QCDT0Rev} for $\varphi = \pi$. The reason is that for $\beta \to \infty$ the boundary conditions to the fields become irrelevant. In contrast, the high temperature limit is highly sensitive to the phase $\varphi$ (see also section \ref{Abschn: Numerik} below). For the usual fermionic boundary conditions, $\varphi = \pi$, also the remaining sum in eq.~(\ref{Gl: ChiralesKondensat1}) can be analytically evaluated  yielding 
\beq
\langle \bar{\psi}^m(\vx) \psi^m(\vx) \rangle_{\varphi = \pi} = -\frac{2 \Ncc}{\pi} \frac{1}{\beta} \int\limits_0^{\infty} \dd p \, \frac{p M(p)}{E(p)} \left\lfloor \frac{\beta p + \pi}{2 \pi} \right\rfloor \, . \label{Gl: ChiralesKondensat2}
\eeq
This expression vanishes for $\beta \to 0$,
\beq
\lim_{\beta \to 0} \langle \bar{\psi}^m(\vx) \psi^m(\vx) \rangle \bigr\vert_{\varphi = \pi} = 0 \, , \label{Gl: ChiralesKondensatHT}
\eeq
which implies the restoration of chiral symmetry in the high temperature phase.

\subsection{Dual condensates}

Inserting the quark condensate  for the phase dependent boundary condition, eq.~(\ref{Gl: ChiralesKondensat1}), into eq.~(\ref{Gl: DualesKondensat}), the integration over the phase $\varphi$ can be analytically evaluated resulting in the dual condensates
\beq
\Sigma_n = -\frac{\Ncc}{\pi^2} \int\limits_0^{\infty} \dd p \, \frac{p^2 M(p)}{E(p)} \Bigl[\delta_{n 0} + \sinc(n \beta p)\Bigr] \, .
\eeq
In particular, for the dressed Polyakov loop $(n = 1)$ we find
\beq
\Sigma_1 = -\frac{\Ncc}{\pi^2} \int\limits_0^{\infty} \dd p \, \frac{p^2 M(p)}{E(p)} \sinc(\beta p) \, . \label{Gl: DualesKondensat1}
\eeq
Obviously, this quantity vanishes in the zero temperature limit
\beq
\lim_{\beta \to \infty} \Sigma_1 = 0 \, , \label{Gl: DualesKondensat0T}
\eeq
while it reaches a finite limit at high temperature
\beq
\lim_{\beta \to 0} \Sigma_1 = -\frac{\Ncc}{\pi^2} \int\limits_0^{\infty} \dd p \, \frac{p^2 M(p)}{E(p)} \, . \label{Gl: DualesKondensatHT}
\eeq
This behavior is expected for the (dressed) Polyakov loop which is an order parameter for confinement. The temperature behavior of $\Sigma_1$ is opposite to that of the usual quark condensate $\langle \bar{\psi} \psi \rangle_{\varphi = \pi}$ which is non-zero at zero temperature but vanishes in the high temperature limit, see eqs.~(\ref{Gl: ChiralesKondensat0T}) and (\ref{Gl: ChiralesKondensatHT}).

\subsection{Critical temperatures}

The explicit expressions, eqs.~(\ref{Gl: ChiralesKondensat1}) and (\ref{Gl: DualesKondensat1}), derived above for the quark condensate and the dressed Polyakov loop show that both quantities depend smoothly on the temperature, i.e.~both the chiral and the deconfinement transitions are realized as crossovers (this point is discussed in the next section). This necessitates a criterion for determining the transition temperatures: In this paper, we will fix them to the inflection points of the quark condensate and dressed Polyakov loop,
\beq
\frac{\partial^2 \langle \bar{\psi}^m(\vx) \psi^m(\vx) \rangle_{\varphi = \pi}}{\partial T^2} \Bigl\vert_{T_{\chi}} = 0 \, , \quad\quad \frac{\partial^2 \Sigma_1}{\partial T^2} \Bigl\vert_{T_{\mathrm{c}}} = 0 \, .
\eeq
From the analytic expressions (\ref{Gl: ChiralesKondensat1}) and (\ref{Gl: DualesKondensat1}), we find
\begin{align}
\frac{\partial^2 \langle \bar{\psi}^m \psi^m \rangle_{\varphi}}{\partial T^2} &= \frac{2 \Ncc}{\pi^2} \beta^4 \int\limits_0^{\infty} \dd p \, \frac{p^4 M(p)}{E(p)} \sum_{l = 1}^{\infty} l^2 \cos(l \varphi) \sinc(l \beta p) \, , \label{Gl: ZweiteAblCK} \\
\frac{\partial^2 \Sigma_1}{\partial T^2} &= \frac{\Ncc}{\pi^2} \beta^4 \int\limits_0^{\infty} \dd p \, \frac{p^4 M(p)}{E(p)} \sinc(\beta p) \, . \label{Gl: ZweiteAblDK}
\end{align}

\section{Numerical results} \label{Abschn: Numerik}

\begin{figure}%
\centering%
\parbox{0.4\linewidth}{%
\centering%
\includegraphics[width=\linewidth]{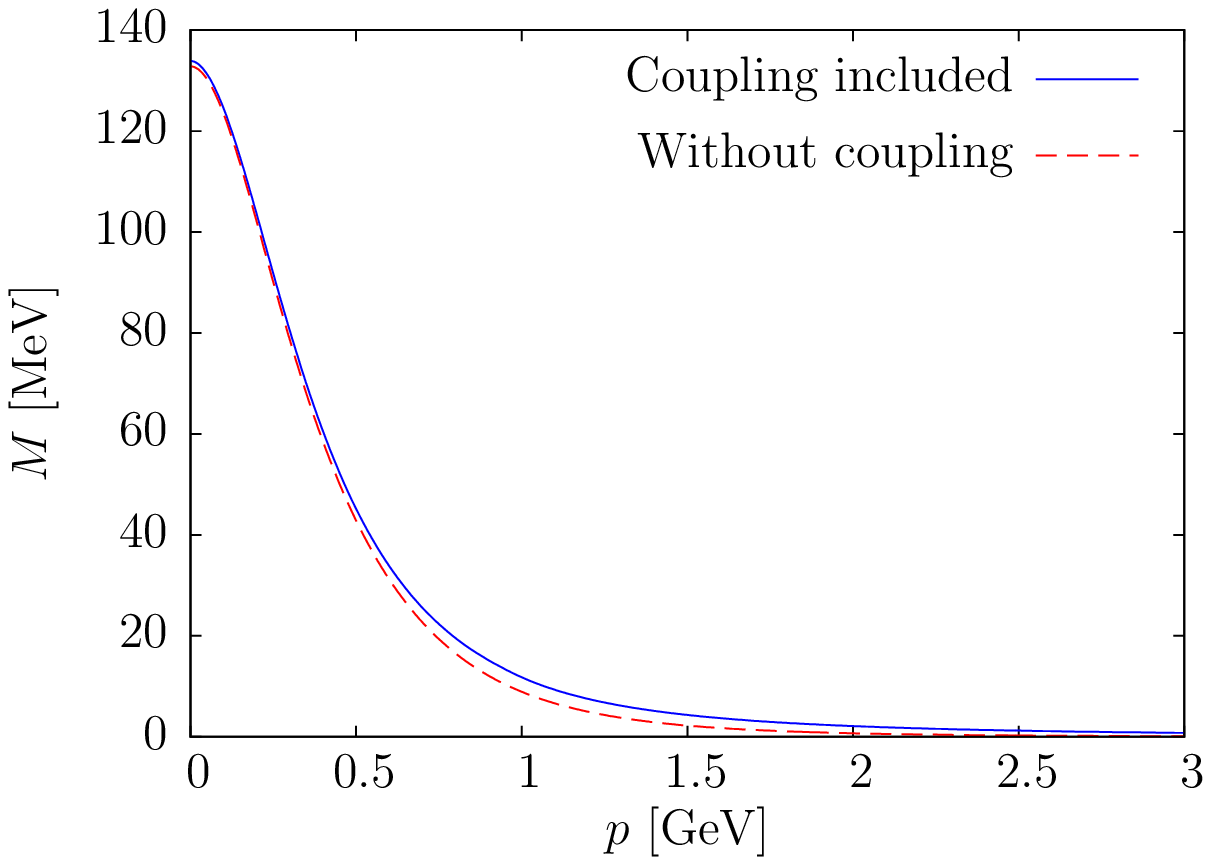} \\%
(a) %
}%
\hspace{0.1\linewidth}%
\parbox{0.4\linewidth}{%
\centering%
\includegraphics[width=\linewidth]{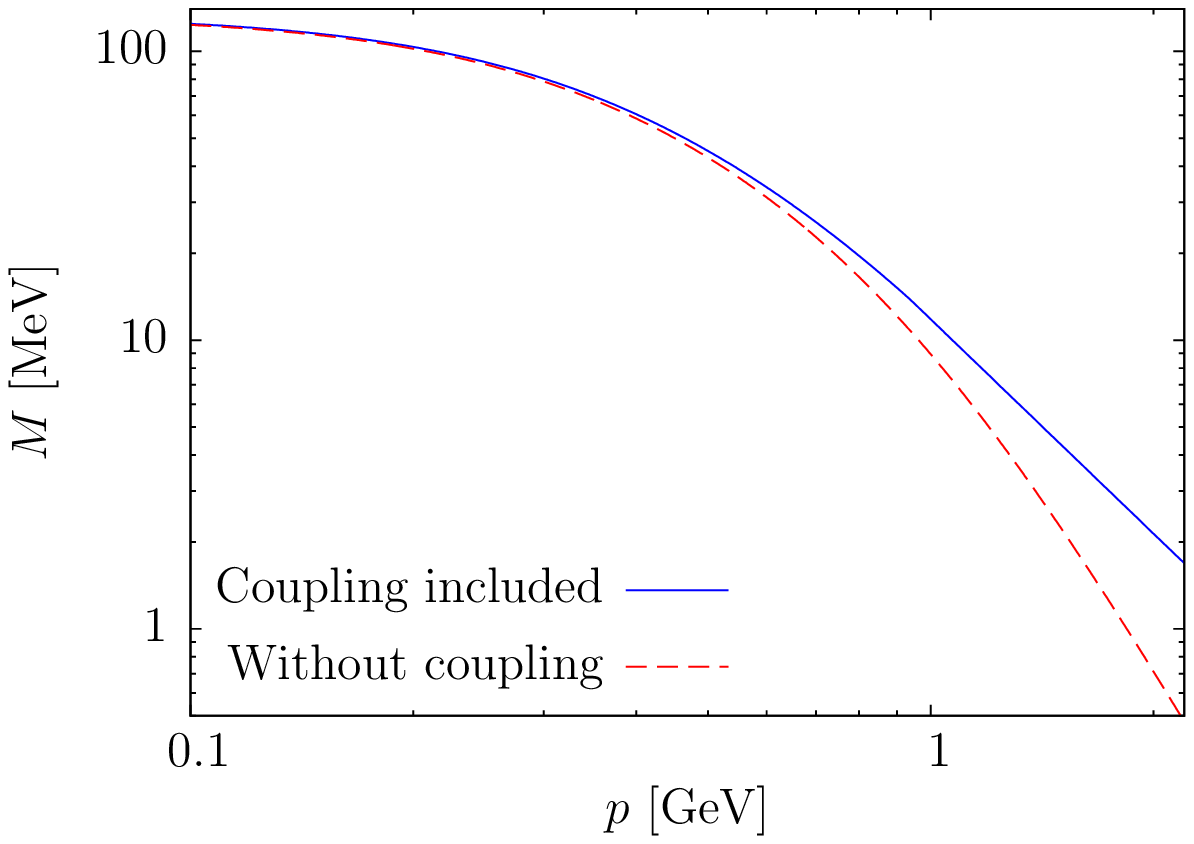} \\%
(b) %
} \\
\parbox{0.4\linewidth}{%
\centering%
\includegraphics[width=\linewidth]{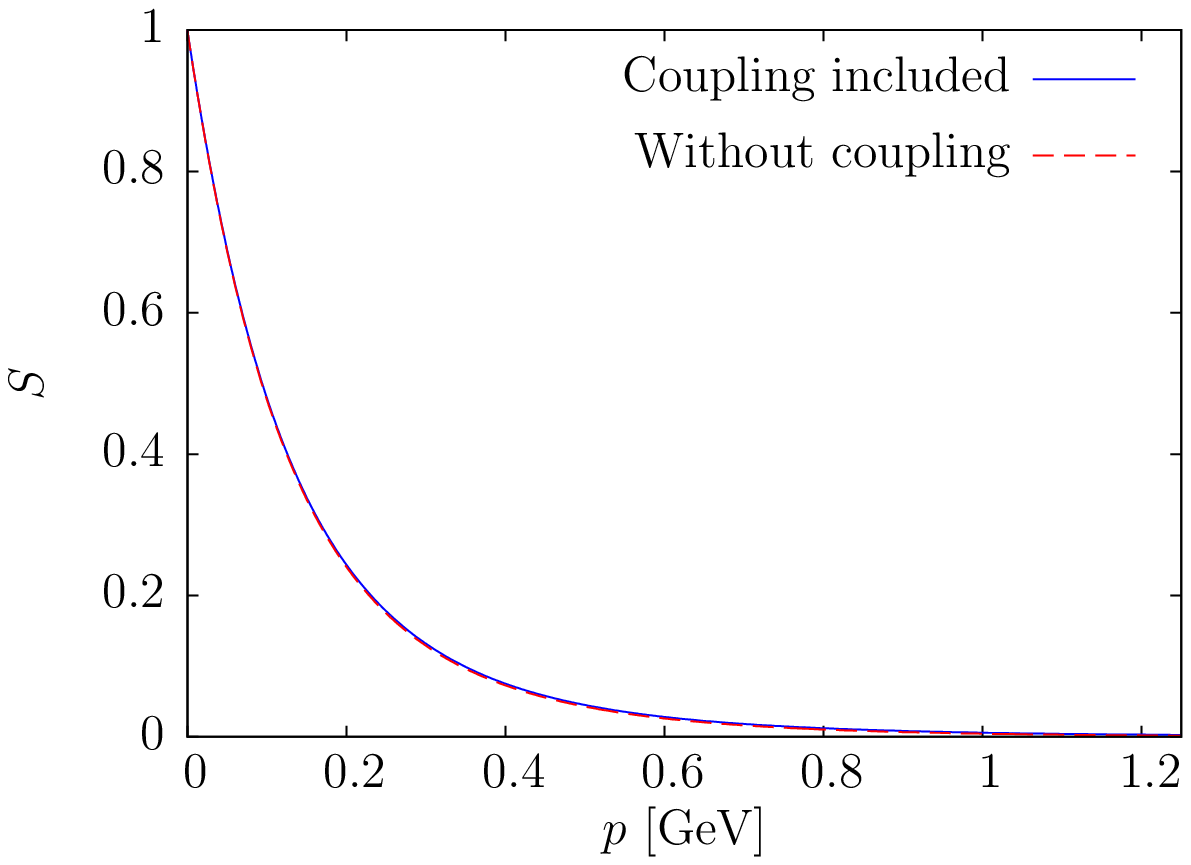} \\%
(c) %
}%
\hspace{0.1\linewidth}%
\parbox{0.4\linewidth}{%
\centering%
\includegraphics[width=\linewidth]{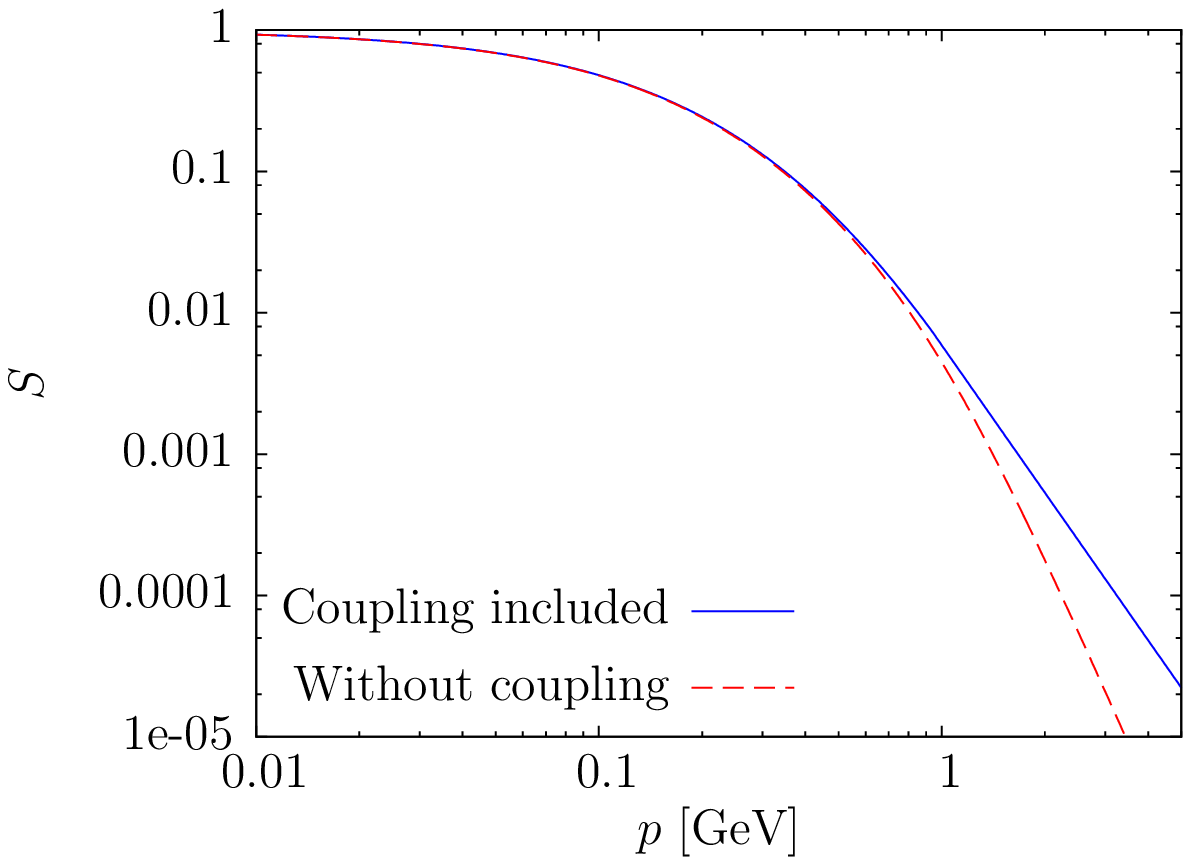} \\%
(d) %
} %
\caption{Zero temperature mass function for a quark-gluon coupling constant of $g = 2.1$ (full curve) compared to the result obtained with a BCS-ansatz in the limit $g = 0$ (dashed curve). Results are presented on both a linear (a) and a logarithmic scale (b). The scalar kernel $S$ is presented in subfigure (c) on a linear and in subfigure (d) on a logarithmic scale.}%
\label{Abb: Massenfunktion}%
\end{figure}

The self-consistent numerical solution of the quark gap equation (\ref{Gl: GapgleichungMassenfunktion}) at finite temperature is quite expensive due to the involved Poisson sums. To reduce the numerical expense in the present paper we use the zero temperature solution to calculate the chiral condensate (\ref{Gl: ChiralesKondensat1}) and the dressed Polyakov loop (\ref{Gl: DualesKondensat1}) and leave the finite temperature solution for future research. In the numerical calculations, the value $\sigma_{\mathrm{C}} = 2.5 \sigma$  \cite{Burgio2012} was assumed for the Coulomb string tension. Furthermore, the quark-gluon coupling $g$ was determined by fixing the chiral quark condensate at zero temperature to its phenomenological value \cite{Williams2007}
\beq
\langle \bar{\psi}(\vx) \psi(\vx) \rangle_{\varphi = \pi}^{\mathrm{phen}} \approx (-235 \, \mathrm{MeV})^3
\eeq
which requires $g \approx 2.1$.\footnote{Within the quenched theory, the quark-gluon coupling constant enters only the quark but not the gluon gap equation. Therefore, $g$ can be fixed within the quark sector.} Figure \ref{Abb: Massenfunktion} shows the solution of the quark gap equation (\ref{Gl: GapgleichungMassenfunktion}) at zero temperature obtained in ref.~\cite{QCDT0Rev}. We present both the scalar variational kernel $S$ and the mass function $M$ [Eq.~(\ref{Gl: Massenfkt})]. For sake of comparison, we also show the results obtained when the coupling of the quarks to the transversal gluons is neglected ($g = 0$),\footnote{Note that here and in the following the limit $g = 0$ always implies the neglecting of both quark-gluon coupling in the wave functional and UV part of the non-Abelian Coulomb potential (\ref{410-f8-1}).} which yields a zero temperature quark condensate of $\langle \bar{\psi} \psi \rangle_{\varphi = \pi} \approx (-185 \, \mathrm{MeV})^3$. As one observes when the coupling and the UV part of the Coulomb potential are included the effective quark mass increases in the mid- and large-momentum regime while in the IR it does practically not change. This is expected since the IR behavior of the gap equation is dominated by the IR part of the Coulomb term which is the only loop term which survives the $g \to 0$ limit of the gap equation, see eqs.~(\ref{Gl: SchleifenintegralMC})-(\ref{Gl: SchleifenintegralMWE}). Note that especially the UV exponent of the mass function significantly increases when the quark-gluon coupling and the UV part of the non-Abelian Coulomb potential (\ref{410-f8-1}) are taken into account, see the logarithmic plot in fig.~\ref{Abb: Massenfunktion}. It is this larger UV exponent which leads to the increase of the quark condensate in the full theory.

\begin{figure}%
\centering%
\parbox{0.4\linewidth}{%
\centering%
\includegraphics[width=\linewidth]{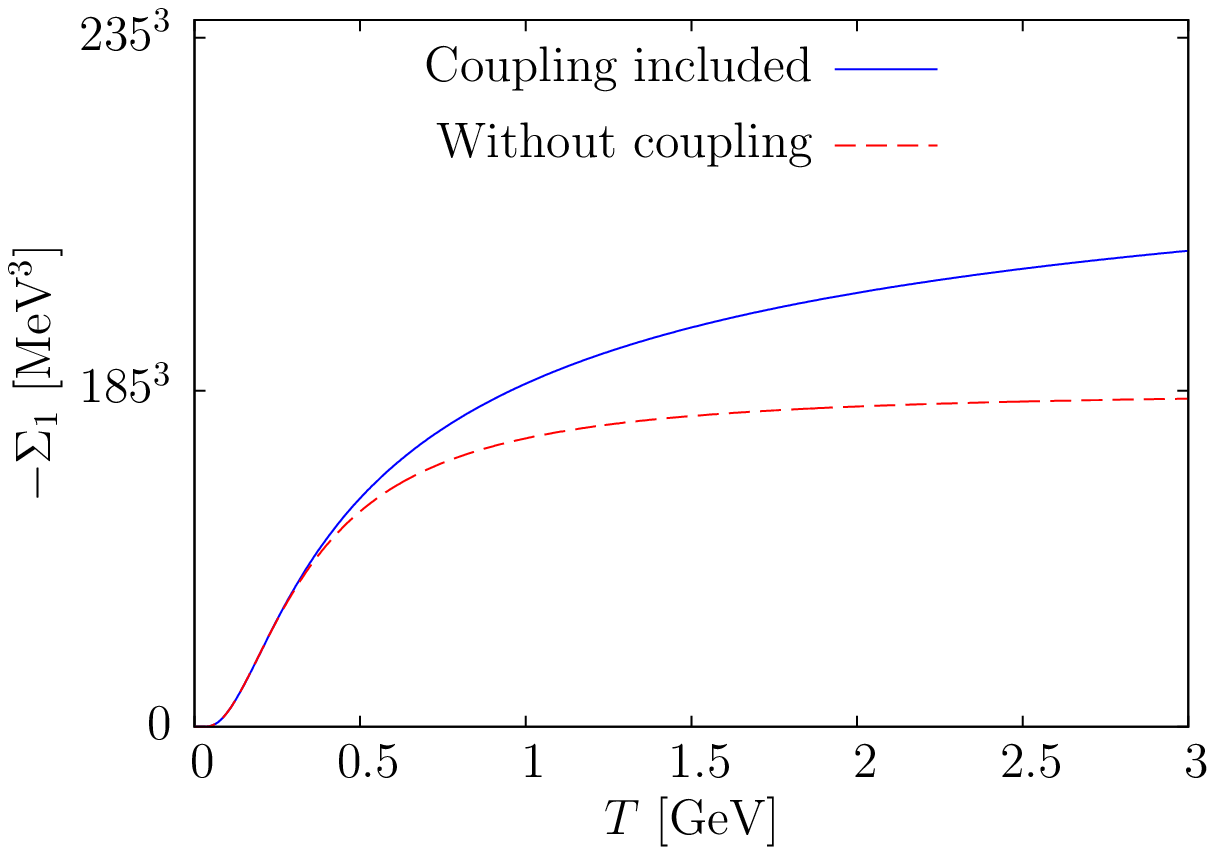} \\%
(a) %
}%
\hspace{0.1\linewidth}%
\parbox{0.4\linewidth}{%
\centering%
\includegraphics[width=\linewidth]{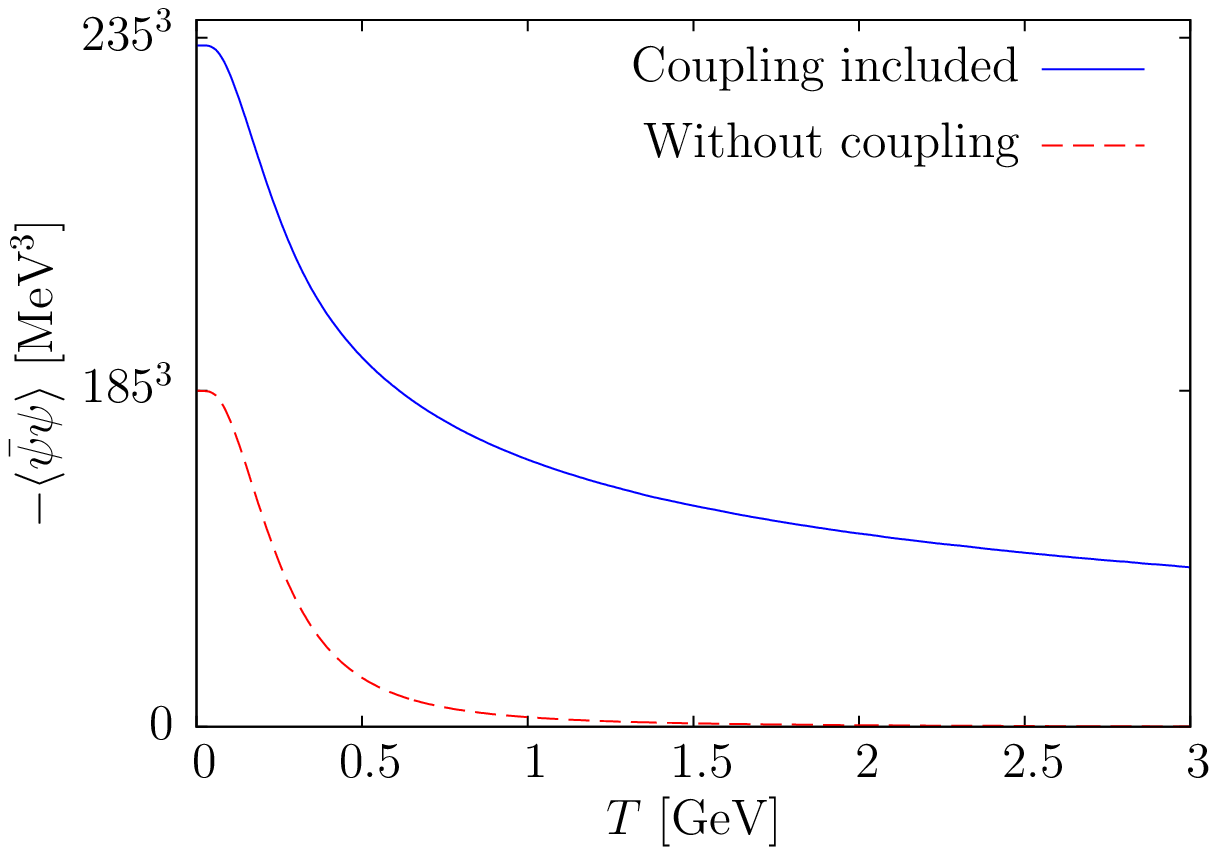} \\%
(b) %
}%
\caption{(a) Dual and (b) chiral quark condensate (with fermionic boundary conditions, eq.~(\ref{680-3}) for $\varphi = \pi$) as a function of the temperature. Both plots show results obtained for $SU(3)$ with the Coulomb string tension set to $\sigma_{\mathrm{C}} = 2.5 \sigma$ and for a quark-gluon coupling constant of $g = 2.1$ (full curve) and $g = 0$ (dashed curve).}%
\label{Abb: DCK}%
\end{figure}

\begin{figure}%
\centering%
\includegraphics[width=0.4\linewidth]{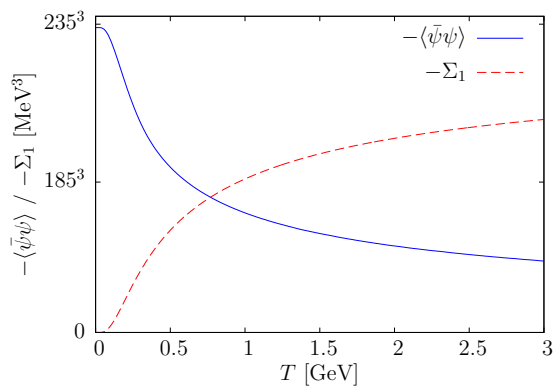} \\%
\caption{Chiral (full curve) compared to the dual quark condensate (dashed curve) as a function of the temperature for $SU(3)$ for a quark-gluon coupling constant of $g = 2.1$.}%
\label{Abb: CDK}%
\end{figure}

In the numerical evaluation of the quark condensates, the Poisson sum was restricted to $l \leq 75$, which turned out to be sufficient for the temperatures considered. The obtained dressed Polyakov loop and chiral quark condensate (for $\varphi = \pi$) are shown in fig.~\ref{Abb: DCK}. For sake of comparison, we also show the results obtained in the limit $g = 0$. As one observes, the inclusion of the coupling of the quarks to the transversal gluons and the UV part of $V_{\mathrm{C}}$ significantly increases both quantities. Figure \ref{Abb: CDK} shows the temperature dependence of the dual and chiral condensate in the same plot. The decrease of the chiral quark condensate is tied to an increase of the dual condensate. Thus, the complementary high and low temperature limits of chiral and dual quark condensate predicted in the last section are confirmed by the numerical results. From these figures it is also seen that both the chiral and the deconfinement transitions are crossovers. On the lattice one finds for the chiral phase transition a crossover for physical (up, down and strange) quark masses while a first order phase transition is observed for sufficiently small quark masses. The latter becomes second order for two chiral quark flavors \cite{Karsch2002}. In our calculation a crossover is obtained even for chiral quarks due to the use of the zero-temperature solution of the gap equation (\ref{Gl: GapgleichungMassenfunktion}) as preliminary investigations show. For the same reason, chiral and dual condensate reach their predicted limits only at very high temperatures, see fig.~\ref{Abb: CDK}.

From the roots of the second derivative of the chiral and dual quark condensate with respect to the temperature, we extract a pseudo-critical temperature of $T_{\mathrm{c}} \approx 196 \, \mathrm{MeV}$ for the deconfinement and $T_{\chi} \approx 168 \, \mathrm{MeV}$ for the chiral transition. Thus, the restoration of chiral symmetry takes place before the deconfinement transition. This result is consistent with lattice measurements. However, in dynamical lattice calculations one finds lower pseudo-critical temperatures of $T_{\mathrm{c}} \approx 165 \, \mathrm{MeV}$ and $T_{\chi} \approx 155 \, \mathrm{MeV}$ for finite quark masses \cite{Borsanyi2010, *Bazavov2012}. In this respect, we should stress that our calculations are not unquenched since we have used the gluon propagator obtained in pure zero temperature Yang--Mills theory. Let us also mention that the pseudo-critical temperatures decrease slightly to $T_{\mathrm{c}} \approx 190 \, \mathrm{MeV}$ and $T_{\chi} \approx 165 \, \mathrm{MeV}$ in the limit $g = 0$, i.e.~when the coupling of the quarks to the transverse gluons and the UV part of the Coulomb potential are neglected.

\begin{figure}
\centering
\includegraphics[width=0.5\linewidth]{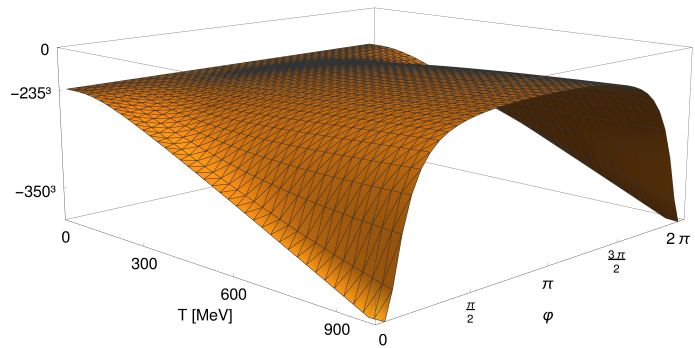}
\caption[]{Chiral quark condensate in units of $\mathrm{MeV}^3$ as a function of the temperature $T$ and the angle $\varphi$ for a quark-gluon coupling constant of $g = 2.1$ and gauge group $SU(3)$.}
\label{Abb: Cphi}
\end{figure}

\begin{figure}
\centering
\includegraphics[width=0.4\linewidth]{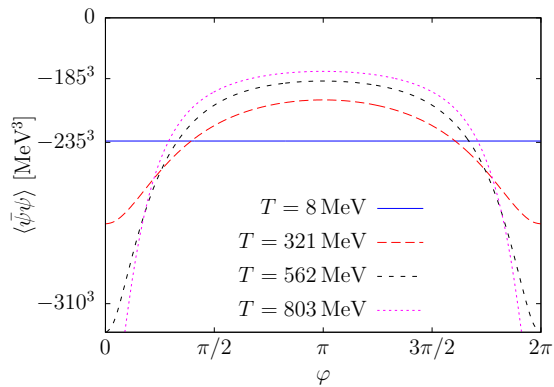}
\caption[]{Chiral condensate as a function of the angle $\varphi$ for different values of the temperature and a quark-gluon coupling constant of $g = 2.1$ and gauge group $SU(3)$.}
\label{Abb: CKphi}
\end{figure}

Figure \ref{Abb: Cphi} shows the quark condensate as function of the temperature and of the angle $\varphi$ of the boundary condition eq.~(\ref{680-3}). As function of the temperature, the magnitude of the condensate $\langle \bar{\psi} \psi \rangle_\varphi$ is decreasing for the usual fermionic boundary condition $(\varphi = \pi)$ and increasing for $\varphi = 0$. Furthermore, the condensate $\langle \bar{\psi} \psi \rangle_\varphi$ is a periodic function of $\varphi$ with period $2 \pi$ and becomes independent of the angle  $\varphi$ for $T \to 0$, in agreement with our analytic observation in section \ref{sectV}. Finally, fig.~\ref{Abb: CKphi} shows the quark condensate as function of the angle $\varphi$ for several temperatures. These curves represent fixed temperature cuts through fig.~\ref{Abb: Cphi} and demonstrate the periodic behavior of the quark condensate. The temperature and boundary angle dependence of the quark condensate $\langle \bar{\psi} \psi \rangle_{\varphi}$ shown in figs.~\ref{Abb: Cphi} and \ref{Abb: CKphi} is in  qualitative agreement with both the lattice data \cite{Zhang2011} and the results of calculations using Dyson--Schwinger equations \cite{Fischer2009, FM2009}.

\section{Summary and Conclusions}\label{sectVII}

In this paper, we have studied QCD at finite temperature within the variational approach in Coulomb gauge developed in ref.~\cite{Feuchter2004, *Feuchter2004a, *Feuchter2005} for the Yang--Mills sector and extended in refs.~\cite{Pak2013, *Pak2012a, QCDT0, QCDT0Rev} to full QCD. The temperature was introduced by compactifying a spatial dimension \cite{Reinhardt2016}. In the finite temperature calculations, we have used the zero temperature variational result for the propagators obtained in refs.~\cite{Feuchter2004, *Feuchter2004a, *Feuchter2005, *ERS2007, QCDT0Rev} as input. With these propagators, we have then calculated the chiral and dual quark condensate as function of the temperature and found in both cases a crossover transition. The pseudo-critical temperatures of the chiral and deconfinement crossover transitions were obtained as $T_{\chi} = 168 \, \mathrm{MeV}$ and $T_{\mathrm{c}} = 196 \, \mathrm{MeV}$. These values are somewhat too large compared to the lattice results. The reason may be that: i) We have used the $T = 0$ variational solution instead of the finite temperature one and ii) we have ignored unquenching effects. Remarkably, one finds somewhat better agreement with the lattice result when one neglects the UV part of the non-Abelian Coulomb potential and the direct coupling of the quarks to the transverse gluon field in the vacuum wave functional. However, in that case, the zero temperature chiral quark condensate is much too small. To improve the present calculations we have first to find the variational finite temperature solutions of the quark sector solving the finite temperature gap equation (\ref{Gl: GapgleichungMassenfunktion}). The unquenching of the gluon propagator has likely a minor effect on the pseudo-critical temperatures while the temperature dependence in the variational solution will certainly have a major effect.

\section*{Acknowledgments}

Discussions with D.~Campagnari and M.~Quandt are greatly acknowledged. This work was supported in part by Deutsche Forschungsgemeinschaft under contract DFG-RE856/9-2.

\begin{appendix}

\section{Momentum representation} \label{Anh: Impulsdarstellung}

Below, we collect the relevant equations of the Hamiltonian approach to QCD in Coulomb gauge in momentum space which are needed for the numerical calculation.

It will be convenient to expand the quark field $\psi$ in terms of the eigenspinors $u$, $v$ of the free Dirac operator of chiral quarks,
\beq
h_0(\vp) = \vec{\alpha} \cdot \vp \, ,
\eeq
which satisfy the eigenvalue equations
\beq
h_0(\vp) \, u^s(\vp) = p u^s(\vp), \quad \quad h_0(\vp) \, v^s(-\vp) = -p v^s(-\vp) \, .
\eeq
Here, $s = \pm 1$ is the double of the spin projection and we choose the following normalization for the spinors
\begin{subequations}
\begin{align}
{u^s}^{\dagger}(\vp) u^t(\vp) &= {v^s}^{\dagger}(-\vp) v^t(-\vp) = 2 p \delta^{s t} \\
{u^s}^{\dagger}(\vp) v^t(-\vp) &= {v^s}^{\dagger}(-\vp) u^t(\vp) = 0 \,.
\end{align}
\label{Gl: SpinorNormierung}%
\end{subequations}
Furthermore, the spinors satisfy
\begin{subequations}
\begin{align}
\gamma_0 u^s(\vp) &= s v^s(-\vp) \\
\gamma_0 v^s(-\vp) &= s u^s(\vp)
\end{align}
\end{subequations}
\begin{subequations}
\begin{align}
\gamma_5 u^s(\vp) &= s u^s(\vp) \\
\gamma_5 v^s(-\vp) &= -s v^s(-\vp)
\end{align}
\end{subequations}
from which we obtain
\begin{subequations}
\begin{align}
\bar{u}^s(\vp) u^t(\vp) &\equiv {u^s}^{\dagger}(\vp) \gamma_0 u^t(\vp) = 0 \\
\bar{v}^s(-\vp) v^t(-\vp) &= 0 \, .
\end{align}
\end{subequations}
For the outer product of the spinors\footnote{Note that there is no summation over the spin index.}
\begin{subequations}
\begin{align}
u^s(\vp) {u^s}^{\dagger}(\vp) &= \frac{1}{2} p \bigl(\left[\Id + \valpha \cdot \hp\right] + s \gamma_5 \left[\Id + \valpha \cdot \hp\right]\bigr) \\
v^s(-\vp) {v^s}^{\dagger}(-\vp) &= \frac{1}{2} p \bigl(\left[\Id - \valpha \cdot \hp\right] - s \gamma_5 \left[\Id - \valpha \cdot \hp\right]\bigr) \\
u^s(\vp) {v^s}^{\dagger}(-\vp) &= \frac{1}{2} p \bigl(s \gamma_0 \left[\Id - \valpha \cdot \hp\right] - \gamma_0 \gamma_5 \left[\Id - \valpha \cdot \hp\right]\bigr) \\
v^s(-\vp) {u^s}^{\dagger}(\vp) &= \frac{1}{2} p \bigl(s \gamma_0 \left[\Id + \valpha \cdot \hp\right] + \gamma_0 \gamma_5 \left[\Id + \valpha \cdot \hp\right]\bigr)
\end{align}
\end{subequations}
holds, while the matrix elements of the free Dirac operator are given by
\begin{subequations}
\begin{align}
{u^s}^{\dagger}(\vp) h_0(\vq) u^t(\vp) &= 2 \vp \cdot \vq \delta^{s t} \\
{v^s}^{\dagger}(-\vp) h_0(\vq) v^t(-\vp) &= -2 \vp \cdot \vq \delta^{s t} \\
{u^s}^{\dagger}(\vp) h_0(\vq) v^s(-\vp) &= 0 \\
{v^s}^{\dagger}(-\vp) h_0(\vq) u^s(\vp) &= 0 \,.
\end{align}
\end{subequations}
The quark field $\psi$ can be expanded as
\beq
\psi^m(\vx) = \int_{\beta} \da^3 p \, \frac{\exp(\ii (\vp_{\perp} + p_{\nf} \he_3) \cdot \vx)}{\sqrt{2} (p_{\perp}^2 + p_{\nf}^2)^{\frac{1}{4}}} \Bigl(a^{s, m}(\vp_{\perp}, p_{\nf}) u^s\bigl(\vp_{\perp} + p_{\nf} \he_3\bigr) + {b^{s, m}}^{\dagger}(-\vp_{\perp}, -p_{\nf}) v^s\bigl(-\vp_{\perp} - p_{\nf} \he_3\bigr)\Bigr) \label{Gl: EntwicklungFeldop}
\eeq
in terms of the eigenspinors $u$, $v$ with $a^s$ ($b^s$) denoting the annihilation operator of a (anti-)quark in a state with spin projection $s/2 = \pm 1/2$. From the normalization (\ref{Gl: SpinorNormierung}) of the eigenspinors, we can conclude that the canonical anti-commutation relations of the fermionic fields $\psi$ are fulfilled, if $a$, $b$ satisfy the relations
\beq
\left\{a^{s, m}(\vp_{\perp}, p_{\nf}), {a^{t, n}}^{\dagger}(\vq_{\perp}, q_{\mf})\right\} = \left\{b^{s, m}(-\vp_{\perp}, -p_{\nf}), {b^{t, n}}^{\dagger}(-\vq_{\perp}, -q_{\mf})\right\} = \delta^{s t} \delta^{m n} \deltabar^2(\vp_{\perp} - \vq_{\perp}) \beta \delta_{\mf, \nf}
\eeq
with all other anti-commutators vanishing. Furthermore, from the expansion (\ref{Gl: EntwicklungFeldop}) we can easily read-off the positive and negative energy components $\psi_{\pm}$ of the quark field,
\beq
\psi_{\pm}(\vx) = \int_{\beta} \dd^3 y \, \Lambda_{\pm}(\vx, \vy) \psi(\vy)
\eeq
with the orthogonal projectors
\beq
\Lambda_{\pm}(\vx, \vy) = \int_{\beta} \da^3 p \, \exp\bigl(\ii (\vp_{\perp} + p_{\nf} \he_3) \cdot(\vx - \vy)\bigr) \Lambda_{\pm}(\vp_{\perp}, p_{\nf}) \, , \quad \quad \Lambda_{\pm}(\vp_{\perp}, p_{\nf}) = \frac{1}{2} \left(\Id \pm \frac{h_0(\vp_{\perp} + p_{\nf} \he_3)}{\sqrt{p_{\perp}^2 + p_{\nf}^2}}\right) .
\eeq
Therefore, our quark vacuum wave functional (\ref{812-11}) acquires the explicit form
\beq
\rvert \phi_{\mathrm{Q}}[A] \rangle = \exp\left(-\int_{\beta} \da^3 p \int_{\beta} \da^3 q \, K^{s t, m n}(\vp_{\perp}, p_{\nf}, -\vq_{\perp}, -q_{\mf}) {a^{s, m}}^{\dagger}(\vp_{\perp}, p_{\nf}) {b^{t, n}}^{\dagger}(-\vq_{\perp}, -q_{\mf})\right) \rvert 0 \rangle
\eeq
in momentum space, where the kernel $K$ is given by
\begin{align}
&K^{s t, m n}(\vp_{\perp}, p_{\nf}, -\vq_{\perp}, -q_{\mf}) = \nonumber \\
&\quad = \delta^{m n} \delta^{s t} \deltabar^2(\vp_{\perp} - \vq_{\perp}) \beta \delta_{\nf, \mf} s S(\vp_{\perp}, p_{\nf}) \nonumber \\
&\quad \phantom{=}\,\, + \frac{g t_a^{m n}}{2 (p_{\perp}^2 + p_{\nf}^2)^{\frac{1}{4}} (q_{\perp}^2 + q_{\mf}^2)^{\frac{1}{4}}} {u^s}^{\dagger}\bigl(\vp_{\perp} + p_{\nf} \he_3\bigr) \Bigl(V(\vp_{\perp}, p_{\nf}, -\vq_{\perp}, -q_{\mf}) + \gamma_0 W(\vp_{\perp}, p_{\nf}, -\vq_{\perp}, -q_{\mf})\Bigr) \nonumber \\
&\quad \phantom{=}\,\, \phantom{+} \times \valpha \cdot \vA^a(\vp_{\perp} - \vq_{\perp}, \omega_{\nf - \mf}) v^t\bigl(-\vq_{\perp} - q_{\mf} \he_3\bigr) \, .
\end{align}
The Fourier transforms of the variational kernels $S$, $V$ and $W$ are assumed to be real valued\footnote{The variational equations obtained within this assumption allow explicitly for a self-consistent solution.} scalar functions and defined by
\begin{align}
S(\vx) &= \int_{\beta} \da^3 p \, \exp\bigl(\ii (\vp_{\perp} + p_{\nf} \he_3) \cdot \vx\bigr) S(\vp_{\perp}, p_{\nf}) \label{Gl: Skalarkern} \\
V(\vx, \vy; \vz) &= \int_{\beta} \da^3 p \int_{\beta} \da^3 q \, \exp\bigl(\ii (\vp_{\perp} + p_{\nf} \he_3) \cdot (\vx - \vz)\bigr) \exp\bigl(-\ii (\vq_{\perp} + q_{\mf} \he_3) \cdot (\vy - \vz)\bigr) V(\vp_{\perp}, p_{\nf}, -\vq_{\perp}, -q_{\mf}) \label{Gl: VKern2} \\
W(\vx, \vy; \vz) &= \int_{\beta} \da^3 p \int_{\beta} \da^3 q \, \exp\bigl(\ii (\vp_{\perp} + p_{\nf} \he_3) \cdot (\vx - \vz)\bigr) \exp\bigl(-\ii (\vq_{\perp} + q_{\mf} \he_3) \cdot (\vy - \vz)\bigr) W(\vp_{\perp}, p_{\nf}, -\vq_{\perp}, -q_{\mf}) \label{Gl: WKern2}
\end{align}
where we have taken into account the total momentum conservation. Note that the scalar kernel $S$ is dimensionless while the vector kernels $V$ and $W$ have dimension of inverse momentum. In order to simplify our calculations, we will assume that the vector kernels fulfill the symmetry relations
\beq
V(\vp_{\perp}, p_{\nf}, -\vq_{\perp}, -q_{\mf}) = V(\vq_{\perp}, q_{\mf}, -\vp_{\perp}, -p_{\nf}) \, , \quad\quad W(\vp_{\perp}, p_{\nf}, -\vq_{\perp}, -q_{\mf}) = W(\vq_{\perp}, q_{\mf}, -\vp_{\perp}, -p_{\nf})
\eeq
and
\beq
V(\vp_{\perp}, p_{\nf}, -\vq_{\perp}, -q_{\mf}) = V(-\vp_{\perp}, -p_{\nf}, \vq_{\perp}, q_{\mf}) \, , \quad\quad W(\vp_{\perp}, p_{\nf}, -\vq_{\perp}, -q_{\mf}) = W(-\vp_{\perp}, -p_{\nf}, \vq_{\perp}, q_{\mf}) \, .
\eeq

\section{Variational equations without Poisson resummation} \label{Anh: Gapgleichung}

Variation of the energy (\ref{Gl: EnergieEW}) with respect to the vector kernels yields two equations which can be solved immediately for $V$, $W$
\begin{align}
V(\vp_{\perp}, p_{\nf}, -\vq_{\perp}, -q_{\mf}) &= \frac{1 + S(\vp_{\perp}, p_{\nf}) S(\vq_{\perp}, q_{\mf})}{N^V(\vp_{\perp}, p_{\nf}, \vq_{\perp}, q_{\mf}) + N^V(\vq_{\perp}, q_{\mf}, \vp_{\perp}, p_{\nf}) + \omega\bigl(|\vp_{\perp} - \vq_{\perp} + (p_{\nf} - q_{\mf}) \he_3|\bigr)} \, , \label{Gl: VKern} \\
W(\vp_{\perp}, p_{\nf}, -\vq_{\perp}, -q_{\mf}) &= \frac{S(\vp_{\perp}, p_{\nf}) + S(\vq_{\perp}, q_{\mf})}{N^W(\vp_{\perp}, p_{\nf}, \vq_{\perp}, q_{\mf}) + N^W(\vq_{\perp}, q_{\mf}, \vp_{\perp}, p_{\nf}) + \omega\bigl(|\vp_{\perp} - \vq_{\perp} + (p_{\nf} - q_{\mf}) \he_3|\bigr)} \label{Gl: WKern}
\end{align}
where we have defined
\begin{align}
N^V(\vp_{\perp}, p_{\nf}, \vq_{\perp}, q_{\mf}) &= \sqrt{p_{\perp}^2 + p_{\nf}^2} P(\vp_{\perp}, p_{\nf}) \Bigl(1 - S^2(\vp_{\perp}, p_{\nf}) + 2 S(\vp_{\perp}, p_{\nf}) S(\vq_{\perp}, q_{\mf})\Bigr) \, , \\
N^W(\vp_{\perp}, p_{\nf}, \vq_{\perp}, q_{\mf}) &= \sqrt{p_{\perp}^2 + p_{\nf}^2} P(\vp_{\perp}, p_{\nf}) \Bigl(1 - S^2(\vp_{\perp}, p_{\nf}) - 2 S(\vp_{\perp}, p_{\nf}) S(\vq_{\perp}, q_{\mf})\Bigr) \, ,  \\
P(\vp_{\perp}, p_{\nf}) &= \frac{1}{1 + S^2(\vp_{\perp}, p_{\nf})} \, .
\end{align}
%Note that, although the two expressions for $V$ and $W$ look quite similar at first sight, their perturbative limit $S \to 0$ is completely different.

For the scalar kernel $S$, variation of the ground state energy (\ref{Gl: EnergieEW}) yields the following gap equation:
\begin{align}
\sqrt{p_{\perp}^2 + p_{\nf}^2} S(\vp_{\perp}, p_{\nf}) &= I_{\mathrm{C}}(\vp_{\perp}, p_{\nf}) + I_{V V}(\vp_{\perp}, p_{\nf}) + I_{W W}(\vp_{\perp}, p_{\nf}) + I_{V \mathrm{Q}}(\vp_{\perp}, p_{\nf}) \nonumber \\
&\phantom{=}\,\, + I_{W \mathrm{Q}}(\vp_{\perp}, p_{\nf}) + I_{V E}(\vp_{\perp}, p_{\nf}) + I_{W E}(\vp_{\perp}, p_{\nf}) \label{Gl: Gapgleichung}
\end{align}
As in the Poisson-resummed equation (\ref{Gl: Gapgleichungresummiert}), the r.h.s.~contains several loop terms which are given by the contribution of the color Coulomb interaction $H_{\mathrm{C}}^{\mathrm{Q}}$ [see eq.~(\ref{802-10})]
\begin{align}
I_{\mathrm{C}}(\vp_{\perp}, p_{\nf}) &= \frac{C_{\mathrm{F}}}{2} \int_{\beta} \da^3 q \, V_{\mathrm{C}}\bigl(|\vq_{\perp} - \vp_{\perp} + (q_{\mf} - p_{\nf}) \he_3|\bigr) P(\vq_{\perp}, q_{\mf}) \nonumber \\
&\phantom{=}\,\, \quad \times \left[S(\vq_{\perp}, q_{\mf}) \bigl(1 - S^2(\vp_{\perp}, p_{\nf})\bigr) - S(\vp_{\perp}, p_{\nf}) \bigl(1 - S^2(\vq_{\perp}, q_{\mf})\bigr) \frac{\vp_{\perp} + p_{\nf} \he_3}{\sqrt{p_{\perp}^2 + p_{\nf}^2}} \cdot \frac{\vq_{\perp} + q_{\mf} \he_3}{\sqrt{q_{\perp}^2 + q_{\mf}^2}}\right] ,
\end{align}
the contribution of the two-loop parts of the free single particle Dirac Hamiltonian $H_{\mathrm{Q}}^0$ (\ref{Gl: DiracHamiltonian})
\begin{align}
I_{V V}(\vp_{\perp}, p_{\nf}) &= -\frac{C_{\mathrm{F}}}{2} g^2 \int_{\beta} \da^3 q \, \frac{V^2(\vq_{\perp}, q_{\mf}, -\vp_{\perp}, -p_{\nf})}{\omega\bigl(\vert \vq_{\perp} - \vp_{\perp} + (q_{\mf} - p_{\nf}) \he_3\vert\bigr)} X\bigl(\vq_{\perp} + q_{\mf} \he_3, -\vp_{\perp} - p_{\nf} \he_3\bigr) P(\vq_{\perp}, q_{\mf}) \nonumber \\
&\phantom{=}\,\, \quad \times \Bigl[\sqrt{q_{\perp}^2 + q_{\mf}^2} P(\vq_{\perp}, q_{\mf}) \Bigl\{S(\vq_{\perp}, q_{\mf}) \Bigl(1 - S^2(\vp_{\perp}, p_{\nf})\Bigr) - S(\vp_{\perp}, p_{\nf}) \Bigl(1 - S^2(\vq_{\perp}, q_{\mf})\Bigr)\Bigr\} \nonumber \\
&\phantom{=}\,\, \quad \phantom{\times \Bigl[} + \sqrt{p_{\perp}^2 + p_{\nf}^2} P(\vp_{\perp}, p_{\nf}) \Bigl\{S(\vq_{\perp}, q_{\mf}) \Bigl(1 - 3 S^2(\vp_{\perp}, p_{\nf})\Bigr) - S(\vp_{\perp}, p_{\nf}) \Bigl(3 - S^2(\vp_{\perp}, p_{\nf})\Bigr)\Bigr\}\Bigr] \, , \\
I_{W W}(\vp_{\perp}, p_{\nf}) &= -\frac{C_{\mathrm{F}}}{2} g^2 \int_{\beta} \da^3 q \, \frac{W^2(\vq_{\perp}, q_{\mf}, -\vp_{\perp}, -p_{\nf})}{\omega\bigl(\vert \vq_{\perp} - \vp_{\perp} + (q_{\mf} - p_{\nf}) \he_3\vert\bigr)} Y\bigl(\vq_{\perp} + q_{\mf} \he_3, -\vp_{\perp} - p_{\nf} \he_3\bigr) P(\vq_{\perp}, q_{\mf}) \nonumber \\
&\phantom{=}\,\, \quad \times \Bigl[\sqrt{q_{\perp}^2 + q_{\mf}^2} P(\vq_{\perp}, q_{\mf}) \Bigl\{S(\vq_{\perp}, q_{\mf}) \Bigl(-1 + S^2(\vp_{\perp}, p_{\nf})\Bigr) - S(\vp_{\perp}, p_{\nf}) \Bigl(1 - S^2(\vq_{\perp}, q_{\mf})\Bigr)\Bigr\} \nonumber \\
&\phantom{=}\,\, \quad \phantom{\times \Bigl[} + \sqrt{p_{\perp}^2 + p_{\nf}^2} P(\vp_{\perp}, p_{\nf}) \Bigl\{S(\vq_{\perp}, q_{\mf}) \Bigl(-1 + 3 S^2(\vp_{\perp}, p_{\nf})\Bigr) - S(\vp_{\perp}, p_{\nf}) \Bigl(3 - S^2(\vp_{\perp}, p_{\nf})\Bigr)\Bigr\}\Bigr] \, ,
\end{align}
the contribution of the quark-gluon coupling in the single particle Dirac-Hamiltonian $H_{\mathrm{Q}}^A$ (\ref{Gl: DiracHamiltonian})
\begin{align}
I_{V \mathrm{Q}}(\vp_{\perp}, p_{\nf}) &= \frac{C_{\mathrm{F}}}{2} g^2 \int_{\beta}\da^3 q \, \frac{V(\vq_{\perp}, q_{\mf}, -\vp_{\perp}, -p_{\nf})}{\omega\bigl(\vert \vq_{\perp} - \vp_{\perp} + (q_{\mf} - p_{\nf}) \he_3\vert\bigr)} X\bigl(\vq_{\perp} + q_{\mf} \he_3, -\vp_{\perp} - p_{\nf} \he_3\bigr) P(\vq_{\perp}, q_{\mf}) \nonumber \\
&\phantom{=}\,\, \phantom{\frac{C_{\mathrm{F}}}{2} g^2 \int_{\beta}\da^3 q \,} \times \Bigl[S(\vq_{\perp}, q_{\mf}) - 2 S(\vp_{\perp}, p_{\nf}) - S(\vq_{\perp}, q_{\mf}) S^2(\vp_{\perp}, p_{\nf})\Bigr] \, , \\
I_{W \mathrm{Q}}(\vp_{\perp}, p_{\nf}) &= \frac{C_{\mathrm{F}}}{2} g^2 \int_{\beta}\da^3 q \, \frac{W(\vq_{\perp}, q_{\mf}, -\vp_{\perp}, -p_{\nf})}{\omega\bigl(\vert \vq_{\perp} - \vp_{\perp} + (q_{\mf} - p_{\nf}) \he_3\vert\bigr)} Y\bigl(\vq_{\perp} + q_{\mf} \he_3, -\vp_{\perp} - p_{\nf} \he_3\bigr) P(\vq_{\perp}, q_{\mf}) \nonumber \\
&\phantom{=}\,\, \phantom{\frac{C_{\mathrm{F}}}{2} g^2 \int_{\beta}\da^3 q \,} \times \Bigl[1 - 2 S(\vq_{\perp}, q_{\mf}) S(\vp_{\perp}, p_{\nf}) - S^2(\vp_{\perp}, p_{\nf})\Bigr]
\end{align}
and the contribution of the kinetic energy of the gluons [Eq.~(\ref{740-2})]
\begin{align}
I_{V E}(\vp_{\perp}, p_{\nf}) &= \frac{C_{\mathrm{F}}}{2} g^2 S(\vp_{\perp}, p_{\nf}) \int_{\beta} \da^3 q \, V^2(\vq_{\perp}, q_{\mf}, -\vp_{\perp}, -p_{\nf}) X\bigl(\vq_{\perp} + q_{\mf} \he_3, -\vp_{\perp} - p_{\nf} \he_3\bigr) P(\vq_{\perp}, q_{\mf}) \, , \\
I_{W E}(\vp_{\perp}, p_{\nf}) &= \frac{C_{\mathrm{F}}}{2} g^2 S(\vp_{\perp}, p_{\nf}) \int_{\beta} \da^3 q \, W^2(\vq_{\perp}, q_{\mf}, -\vp_{\perp}, -p_{\nf}) Y\bigl(\vq_{\perp} + q_{\mf} \he_3, -\vp_{\perp} - p_{\nf} \he_3\bigr) P(\vq_{\perp}, q_{\mf}) \, .
\end{align}
Here we used the definition of $X$ and $Y$ given in eq.~(\ref{Gl: Winkelfaktoren}). After Poisson resummation (\ref{305-f2a-x1}) of the above given equations, one arrives at the variational equations given in section \ref{subsecC}.

\end{appendix}

\bibliography{DualesKondensat}

\end{document}